\documentclass{aa}  

\usepackage{graphicx}
\usepackage{txfonts}
\usepackage{array}
\usepackage{tabularx}
\usepackage{amsmath}
\usepackage{multirow}
\usepackage{multicol}
\usepackage{booktabs} 
\usepackage{natbib}
\bibpunct{(}{)}{;}{a}{}{,} 

\usepackage{comment}
\usepackage{color}
\usepackage[normalem]{ulem}

\newcommand{\msun}{M$_{\sun}$}

\newcommand{\net}{\emph{net90}}
\newcommand{\netnoec}{\emph{net90-noec}}
\newcommand{\neteightsix}{\emph{net86}}
\newcommand{\netalpha}{\emph{net14}}
\newcommand{\netWN}{\emph{WinNet}}
\newcommand{\netWNnoec}{\emph{WinNet-noec}}
\newcommand{\dens}{~g$\cdot$cm$^{-3}$}
\newcommand{\carbon}{$^{12}$C}
\newcommand{\oxygen}{$^{16}$O}
\newcommand{\ele}{$e^-$}
\newcommand{\pos}{$e^+$}
\begin{document}

   \title{Do not forget the electrons: Extending moderately-sized nuclear networks for multidimensional hydrodynamic codes}
   \titlerunning{Do not forget the electrons: Extending nuclear networks for multi-D hydrodynamic codes}

   \author{Domingo García-Senz\inst{1,2}
          \and
          Rubén M. Cabezón\inst{3}
          \and
          Moritz Reichert\inst{4}
          \and
          Axel S. Lechuga\inst{1}
          \and
          José A. Escartín\inst{3}
          \and 
          Athanasios Psaltis\inst{5,6}
          \and
          Almudena Arcones\inst{7,8,9}
          \and
          Friedrich-Karl Thielemann\inst{8,10}
            }

   \institute{Departament de Física. Universitat Politècnica de Catalunya (UPC). Av. Eduard Maristany 16, 08019 Barcelona, Spain\\
    \email{domingo.garcia@upc.edu}
    \and
    Institut d'Estudis Espacials de Catalunya (IEEC), 08860 Castelldefels (Barcelona), Spain
    \and
    Center for Scientific Computing - sciCORE, University of Basel, Klingelberstrasse 61, 4056 Basel, Switzerland\\
    \email{ruben.cabezon@unibas.ch}
    \and 
    Department of Astronomy and Astrophysics, University of Valencia, C/Dr. Moliner 50, E-46100 Burjassot (Valencia), Spain\\
    \email{moritz.reichert@uv.es}
    \and
     Triangle Universities Nuclear Laboratory, Duke University, Durham, North Carolina 27708, USA
     \and
   Department of Physics, Duke University, Durham, North Carolina 27708, USA
     \and
    Institut für Kernphysik, Technische Universität Darmstadt, Schlossgartenstr. 2, Darmstadt 64289, Germany
    \and
    GSI Helmholtzzentrum für Schwerionenforschung GmbH, PlanckStrs. 1, Darmstadt 64291, Germany
    \and
    Max-Planck-Institut für Kernphysik, Saupfercheckweg 1, 69117 Heidelberg, Germany
    \and
    Department of Physics, University of Basel, Klingelbergstrasse 82, 4056 Basel, Switzerland
    }

   \date{}

 
  \abstract
   {Nuclear networks are widely used coupled with hydrodynamical simulations of explosive scenarios to account for the change of nuclear species and energy generation rate due to nuclear reactions. In this way, there is a feedback mechanism between the hydrodynamical state and the nuclear processes. Unfortunately, the timescale of nuclear reactions is orders of magnitude smaller than the dynamical timescale that drives hydrodynamical simulations. Therefore, these nuclear networks are usually very small, reduced in most cases to a dozen elements, especially when simulations are carried out in more than one dimension.}
   {We present here an extended nuclear network, with 90 species, designed for being coupled with hydrodynamic simulations, which includes neutrons, protons, electrons, positrons, and the corresponding neutrino and anti-neutrino emission. This network is also coupled with temperature, making it extremely robust and, together with its size, unique of its kind. The inclusion of electron captures on free protons makes the network very appropriate for multidimensional studies of Type Ia supernova explosions, especially when the exploding object is a massive white dwarf.}
   {We perform several tests that are relevant to simulate explosive scenarios, such as Type Ia supernovae and core-collapse supernovae. We compare the results of the 90 nuclei network with a standard $\alpha$-chain network with 14 elements to evaluate the differences in the energy generation rate. We also evaluate the relevance of including the electrons in the network in terms of generated yields and how it affects the pressure of a degenerate fluid such as that of white dwarfs. The results obtained with the 90-nuclei network have been verified with a much larger 2000-nuclei network built from REACLIB (WinNet), in terms of nuclear energy generation rate, pressure, and produced yields.}
   {The results obtained with the proposed medium-sized network compare fairly well, to a few percent, with those computed with \netWN{} in scenarios reproducing the gross physical conditions of current Type Ia supernova explosion models. In those cases where the carbon and oxygen fuel ignites at high density, the high-temperature plateau typical of the nuclear statistical equilibrium regime is well defined and stable, allowing large integration time steps. We show that the inclusion of electron captures on free protons substantially improves the estimation of the electron fraction of the mixture. Therefore, the pressure is better determined than in networks where electron captures are excluded, which will ultimately lead to more reliable hydrodynamic models. Explosive combustion of helium at low density, occurring near the surface layer of a white dwarf, is also better described with the proposed network, which gives nuclear energy generation rates much closer to \netWN{} than typical reduced alpha networks. }
   {A nuclear network with N=90 species, including electrons, aimed at multidimensional calculations of supernova explosions is described and verified. The proposed network is suitable for the study of Type Ia supernova explosions because it provides better values of pressure and electron abundance than other existing networks with smaller or even a similar size but without including electron capture processes.}

   \keywords{Methods: numerical: - Nuclear reactions --
                supernova -- nucleosynthesis 
               }

   \maketitle
%

\section{Introduction}

Understanding cosmic violent phenomena such as novae or both types of supernova explosions demands not only a good knowledge of nuclear reaction cross sections but a careful implementation of the nuclear networks in the hydrodynamic codes aimed at simulating these explosions. This issue is especially relevant in the case of Type Ia supernova explosions (SNIa), where the disrupting mechanism is of thermonuclear origin and strong feedback between the hydrodynamics and the released nuclear energy is expected \citep{nom84, nie00}. 

The question of the optimal size of these nuclear networks has always challenged the community and is, among other things, closely connected to the available computational resources. Nowadays, hydrodynamic codes that simulate SNIa explosions are, for the most part, of multidimensional nature. The number of spatial and temporal steps involved in the simulation of one of these explosions is huge, which poses practical limits to the size of nuclear networks. Pioneering multidimensional simulations of SNIa incorporated around 10 nuclei in an attempt to have, at best, a reasonable depiction of the released nuclear energy. For example, \cite{benz1989} implemented an $\alpha$ network to keep track of the nuclear evolution during the collision of two white dwarfs. \cite{Timmes2000} analyzed the performance of two networks, with 7 and 14 species, and concluded that these can account for the nuclear energy generation rate within a $20\%$ precision on average. The use of these small networks has been commonplace in multidimensional simulations of supernova explosions around the edge of the last century, as \cite{Reinecke99, gar99, Plewa2004}, to mention a few connected with different SNIa explosion scenarios. Currently, multidimensional simulations incorporate several dozens of nuclei, which can reproduce the released nuclear energy within a narrow deviation, typically a few percent on average when compared to a larger network \citep{town2019,gronow21}.

However, with current computing technology and numerical methods, it is nowadays feasible to extend the number of nuclei in the network to around one hundred \citep[e.g.,][]{Harris2017,Sandoval2021,navo2023}. In that case, the matching of the released nuclear energy and other magnitudes compared to the post-processed values would benefit further. Nevertheless, the post-processing step would still be necessary to obtain nucleosynthetic details of the produced yields.       

In this work, we propose a network of 90 species, \net{}\footnote{\url{https://github.com/rmcabezon/net90}}, which can be used to model all kinds of SNIa explosions. As we will see, it can be used to assist in the explosion of a massive white dwarf (WD) made of $^{12}$C$+^{16}$O with central density $\rho_c\simeq 2\times 10^9$~\dens{} in the so-called Chandrasekhar-mass explosion models. It is also adequate to track the detonation of a tiny helium shell at densities $\simeq 5\times 10^5$~\dens{} located on top of a moderately massive WD, which is appropriate for studying the Sub-Chandrasekhar mass route to SNIa. \net{} also improves over existing reduced, $\alpha$-like, networks regarding core collapse supernova (CCSN) scenarios, but its use is restricted to the stages with low or moderate neutronization.

Our proposal has two important features worth mentioning. First, the abundances of the species $Y_i=X_i/A_i$ (where $X_i$ is the mass fraction and $A_i$ the atomic mass of the species $i$) and the temperature $T$ are found jointly, after implicitly solving the system of nuclear reactions coupled with the energy equation. It is well known that the implicit coupling of $\{Y_i, T\}$ leads to a very stable behavior around the nuclear statistical equilibrium (NSE) regime; \cite{mueller86}, \cite{cab04} (paper I afterward). The implicit coupling between released nuclear energy, nuclear reaction rates, and temperature makes the scheme robust in simulating high-density combustion, allowing it to naturally handle the quasi (QNSE) and complete nuclear statistical equilibrium (NSE) stages smoothly, without switching to specific NSE routines\footnote{Switching to a specific NSE routine (or a Table) introduces extra parameters in the calculation, for example to decide when the nuclear network is switched-off/on to enter/leave the NSE}. It also makes unnecessary the implementation of burning limiters for the simulation of detonation waves \citep{zin21}. 

The second novelty of this work is that our moderately-sized network includes the capture reactions $e^- + p \rightarrow n + \nu_e$~and $e^{+}+ n \rightarrow p + \bar\nu$. As far as we know, this is the first time that reactions of this kind have been taken into account in a reduced network routine belonging to a multi-D hydrodynamic code dealing with SNIa explosions. As we show below, the benefits of including this reaction, namely a more accurate depiction of the electron pressure and neutronized yields, largely compensate for the shortcoming of a tiny increase ($<0.5\%$) in computing time.      

In Sect.~\ref{sec:formalism} we summarize the mathematical formalism used to integrate the chemical and temperature equations. The main features of the different nuclear networks used in this work are described in Sect.~\ref{sec:nuclear_net}. Section~\ref{sec:tests} is devoted to verifying \net{} and comparing it with a much larger nuclear network in several scenarios. The computational performance of \net{} and related issues are discussed in Sect.~\ref{sec:performance}. Finally, we provide a summary and discussion of the results and prospects for the future in Sect.~\ref{sec:conclusions}.

\section{Mathematical formalism}
\label{sec:formalism}

We follow the method described in paper~I and in \cite{sanz2022} to integrate the nuclear equations. We summarize here the main features of the method. 

The abundances $Y_i$ of the species evolve according to the set of ordinary differential equations, 

\begin{equation}
 \frac{dY_i}{dt}=\sum_{k,l}{r_{kl}Y_lY_k}-\sum_{j}{r_{ij}Y_iY_j}+\sum_{m}{\lambda_mY_m}-\lambda_iY_i\,.
 \label{Eq:nuceq_1}
\end{equation}

The magnitudes $r_{ij}=r_{ij}(\rho,T)=\rho N_A \langle \sigma,\nu \rangle_{ij}$ correspond to the nuclear reaction rates, and $\lambda_i=\lambda_{i}(T)$ stands for $\gamma$ photodisintegration rates, $\beta^+$ decays, and/or $e^{\pm}$ capture processes.

Following paper~I, the energy equation can be written as,  

\begin{equation}
\small
\begin{aligned}
dQ&+\sum_i{\left({BE}_i-\frac{\partial U}{\partial Y_i}\right) dY_i}+ \left(B_{pn} -\frac{\partial U}{\partial Y_e}\right) dY_e-\left(\frac{\partial U}{\partial T}-\frac{d\rho}{\rho^2}\frac{\partial P}{\partial T}\right)dT\\
&=-T\frac{d\rho}{\rho^2}\left(\frac{\partial P}{\partial T}\right)\,,
\end{aligned}
\label{Eq:nuceq_2}
\end{equation}

\noindent where $BE$ is the nuclear binding energy of the nuclei and, in this work, $dQ\equiv-\dot\nu~dt$ accounts for the neutrino losses, $U$ and $P$ are the internal energy and pressure of the gas, and $B_{pn}=0.782$~MeV is the proton+electron to neutron mass-difference. 
This set of equations is usually integrated keeping $\{\rho, T\}$ constant during the current timestep $\Delta t$. Once the new abundances are known, the nuclear energy released in $\Delta t$ is obtained and added to the energy equation. However, this procedure leads to problems during explosive high-temperature combustion, where the rates of direct and inverse reactions are very large, yet they may balance. Close to or at NSE the integration becomes unstable, demanding very small timesteps or even making the calculation non-feasible. However, the joint implicit integration of the abundance and energy equations has been shown to restore the tractability of the problem \citep{mueller86}.  

We first write the set of chemical equations (Eq.~\ref{Eq:nuceq_1}) plus the energy equation (Eq.~\ref{Eq:nuceq_2}) in a suitable form to obtain an implicit solution via a Newton-Raphson (NR) scheme. For example, Eq.~\ref{Eq:nuceq_1} is written as,

\begin{equation}
\label{Eq:nuceq_3}
\begin{split}
 \frac{{Y_i}^{n+1}-{Y_i}^{n}}{\Delta t} &= \sum_{k,l}{r_{kl}(T^{n+\theta})Y_l^{n+\theta}Y_k^{n+\theta}}-\sum_{j}{r_{ij}(T^{n+\theta})Y_i^{n+\theta}Y_j^{n+\theta}}\\ 
 &+\sum_{m}{\lambda_m(T^{n+\theta})Y_m^{n+\theta}}-\lambda_i(T^{n+\theta})Y_i^{n+\theta}\,,
\end{split}
\end{equation}

\noindent
where the notation $Y_i^{n+\theta}=Y_i^{n}+\theta\Delta Y_i$ and $T^{n+\theta}=T^{n}+\theta\Delta T$ is used to choose among explicit, implicit, or mixed schemes by varying the value of the parameter $\theta~[0,1]$. For convenience, we use the simplified notation $r_{ij}^{n+\theta}\equiv r_{ij}(T^{n+\theta})$ and $\lambda_i^{n+\theta}\equiv \lambda_i(T^{n+\theta})$, hereafter. Furthermore, we perform a first-order Taylor expansion, $r_{ij}^{n+\theta}\simeq r_{ij}^n+ \theta ~r{'}^n_{ij} \Delta T$, where $r'$ is the derivative of the rate with respect to the temperature (and also for $\lambda^{n+\theta}_i$). We apply the NR iterative scheme to solve the nonlinear system of Eq.~\ref{Eq:nuceq_3} together with the energy equation (Eq.~\ref{Eq:nuceq_2}). At the integration step $n+1$ and NR $k-$iteration, the corrections $\delta Y_i^{n+1}$ to $(Y_i^{n+1})^{k-1}$ and $\delta T^{n+1}$ to $(T^{n+1})^{k-1}$ are found from\footnote{For the sake of clarity we henceforth omit the symbol $k$ indicating the current NR iteration, $(Y_i^{n+1})^{k-1}\equiv Y_i^{n+1}$ and $(T^{n+1})^{k-1}\equiv T^{n+1}$},
\begin{figure*}
\centerline{\includegraphics[width=1\textwidth]{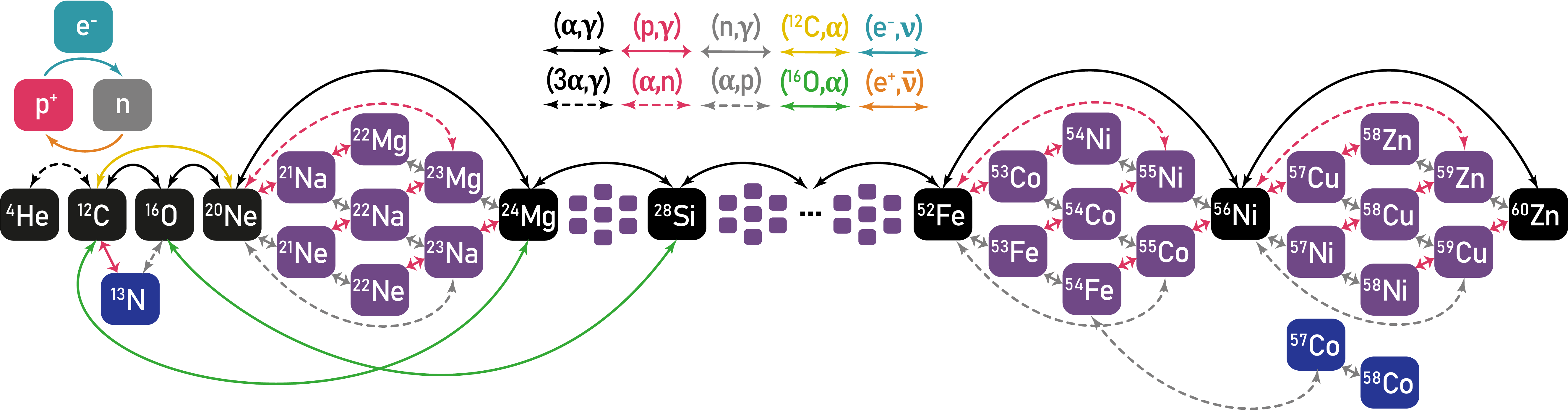}}
\hfil
\caption{Sketch of the architecture of \net{}. There are shown, in blue, the three additional nuclei with respect \neteightsix{} plus electrons (in green)}
\label{fig:arquitecture_net90}
\end{figure*}

\begin{equation}
\begin{split}
 &\frac{\delta Y_i^{n+1}}{\Delta t} = \theta\sum_{k,l}r_{kl}^{n+\theta}Y_l^{n+\theta}\delta Y_k^{n+1}+\theta\sum_{k,l}{r_{kl}^{n+\theta}Y_k^{n+\theta}\delta Y_l^{n+1}}\\
 & -\theta\left(\sum_j{{r_{ij}^{n+\theta}Y_j^{n+\theta}}}\right)\delta Y_i^{n+1}
 -\theta\sum_j{{r_{ij}^{n+\theta}Y_i^{n+\theta}}\delta Y_j^{n+1}}
+\theta\sum_m\lambda_m^{n+\theta}\delta Y_m^{n+1}\\
 &-\theta \lambda_i^{n+\theta}\delta Y_i^{n+1}
+\theta~ \Bigg(\sum_{k,l}{r'}_{kl}^{n+\theta}Y_l^{n+\theta}Y_k^{n+\theta}-\sum_{j}{r'}_{ij}^{n+\theta}Y_i^{n+\theta}Y_j^{n+\theta}\\
 &+\sum_{m}{{\lambda{'}}_m^{n+\theta}Y_m^{n+\theta}}+{\lambda{'}}_i^{n+\theta}Y_i^{n+\theta} \Bigg)~\delta T^{n+1}
 +\sum_{k,l}{r_{kl}^{n+\theta}Y_l^{n+\theta}Y_k^{n+\theta}}\\
&-\sum_{j}{r_{ij}^{n+\theta}Y_i^{n+\theta}Y_j^{n+\theta}}
 +\sum_{m}{\lambda_m^{n+\theta}Y_m^{n+\theta}}-\lambda_i^{n+\theta}Y_i^{n+\theta}-\frac{Y_i^{n+1}-Y_i^n}{\Delta t}\,,
\end{split}
\label{Eq:nuceq_4}
\end{equation}

\noindent
for the chemical equations, and from,

\begin{equation}
\begin{split}
&\sum_k \left(BE_k-\frac{\partial U_{ions}}{\partial Y_k}\right) \delta Y_k^{n+1}+ \left(B_{pn}-\frac{\partial U_e}{\partial{Y_e}}-\frac{\partial (\dot{\nu}\Delta t)}{\partial{Y_e}}\right) \delta Y_e^{n+1}\\
&-\left(\frac{\partial U}{\partial T}-\frac{\Delta\rho}{\rho^2}\frac{\partial P}{\partial T}+\frac{\partial(\dot{\nu}\Delta t)}{\partial T}\right)\delta T^{n+1}= 
\left(\frac{\partial U}{\partial T}-\frac{\Delta\rho}{\rho^2}\frac{\partial P}{\partial T}\right)(T^{n+1}-T^n)\\
&-\sum_k \left(BE_k-\frac{\partial U_{ions}}{\partial Y_k}\right)(Y_k^{n+1}-Y_k^n) -\left(B_{pn}-\frac{\partial U_e}{\partial Y_e}\right)(Y_e^{n+1}-Y_e^n)\\ &+\dot\nu\Delta t-T\frac{\Delta\rho}{\rho^2}\left(\frac{\partial P}{\partial T}\right)\,,
\end{split}
\label{Eq:nuceq_5}
\end{equation}

\noindent
for the energy equation. Equation~\ref{Eq:nuceq_5} includes the neutrino losses, $\dot\nu \Delta t$, produced by electron and positron captures, and the remaining symbols have the usual meaning.

Equations~\ref{Eq:nuceq_4} and \ref{Eq:nuceq_5} conform to an $(N+1)\times (N+1)$ linear system of equations with $N$ unknown chemical corrections, $\delta Y_k$, plus the temperature correction $\delta T$. The system is sparse and is efficiently solved with an improved version of the algorithm described in \cite{prantzos1987}. As an initial guess in the first NR iteration, we take $Y^{n+1}= Y^n; T^{n+1}=T^n$ and the parameter $\theta$ in Eq.~\ref{Eq:nuceq_4} was set to $\theta=0.7$ in the tests described in Sect.~\ref{sec:tests}. This choice of $\theta$ was empirically chosen to balance between $\theta=1$ providing optimal stability and $\theta=0.5$ of perfectly centered integration schemes.    

\section{Nuclear networks}
\label{sec:nuclear_net}
In this section, we introduce a family of nuclear networks, which we named \emph{netX}, where $X$ indicates the number of species. From simpler to the more complex, \emph{net14}, \emph{net86}, and the new \net{} presented here. All of these are based on REACLIB. Verification of \net{} is performed by comparing its results with those obtained with the state-of-the-art nuclear network \emph{WinNet} \citep{reichert2023}, a much larger network that includes many weak interaction processes. The main features of all these nuclear networks are described below.

\subsection{\emph{net14}}
This network was introduced in \cite{gar03} and it is representative of typical $\alpha$-networks, widely used coupled to hydrodynamic codes. This network follows an alpha-chain of reactions from $^4$He up to $^{60}$Zn. Many similar networks had either fewer nuclear species such as \cite{Timmes2000} with only seven nuclei or a bit more, as in \cite{weaver1978} with 19 isotopes. However, the main characteristic of \emph{net14} is that it solves the equations for the evolution of nuclear species coupled with the temperature equation. This methodology was developed with \emph{net14} and used later in the other two networks, \emph{net86} and \net{}.
Small networks such as \emph{net14} are suitable to follow the energy generation rate in explosive scenarios such as Type Ia supernovae, having reasonable feedback with the hydrodynamic state of the fluid. Their accuracy in the energy generation rate is expected to be of the order of $\simeq 20\%$ \citep{Timmes2000} and their nucleosynthetic yields must be post-processed with larger networks in the range of thousands of nuclei. Despite this, their popularity is due to their reduced size, which results in a very small computational footprint, even when they have to be called many times per fluid element and per hydrodynamic timestep.

\subsection{\neteightsix{}}
This network was born as an evolution of \emph{net14}, with the objective of including protons and neutrons, and was also described in \cite{cab04}. To do so, many more nuclear species must be included, and if not careful, the size of the network can increase extremely fast. To that extent, we built this network with a set of blocks of neutron and proton captures around the $\alpha$-chain, again from $^4$He up to $^{60}$Zn. The result is an extended network with 86 nuclei and $\sim 150$ reactions. This is a reasonable compromise between accuracy (for energy generation rate and yields) and computational burden.
\emph{net86} also solves the abundance equations jointly with the temperature equation, which proved to be extremely robust, something that becomes even more important when we increase the number of species.

\begin{figure}
\centerline{\includegraphics[width=0.5\textwidth]{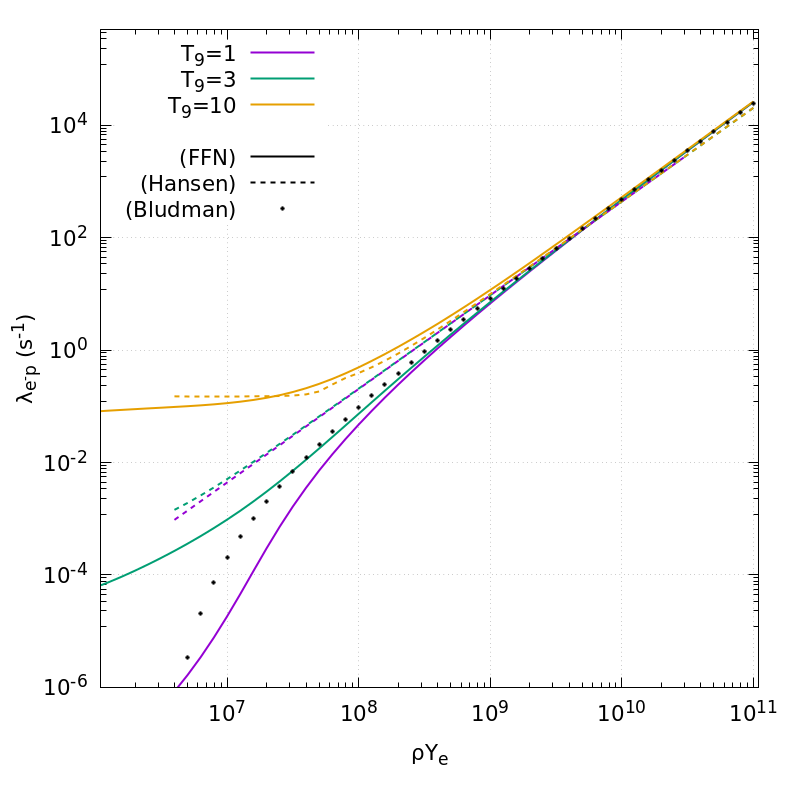}}
\hfil
\caption{Comparison among $e^-$ capture rates at temperatures $T_9=1$, $3$, $10$, and different values of $\rho Y_e$. The rate obtained numerically with the exact formulation of FFN (solid lines, \citealt{Fuller1982}) is compared to those obtained with the analytical approach by \cite{Hansen1968} (dashed lines) and the zero temperature limit rate by \cite{Bludman1982} (black dotted line).}
\label{fig:figratescomp}
\end{figure}

\begin{figure}
\centerline{\includegraphics[width=0.5\textwidth]{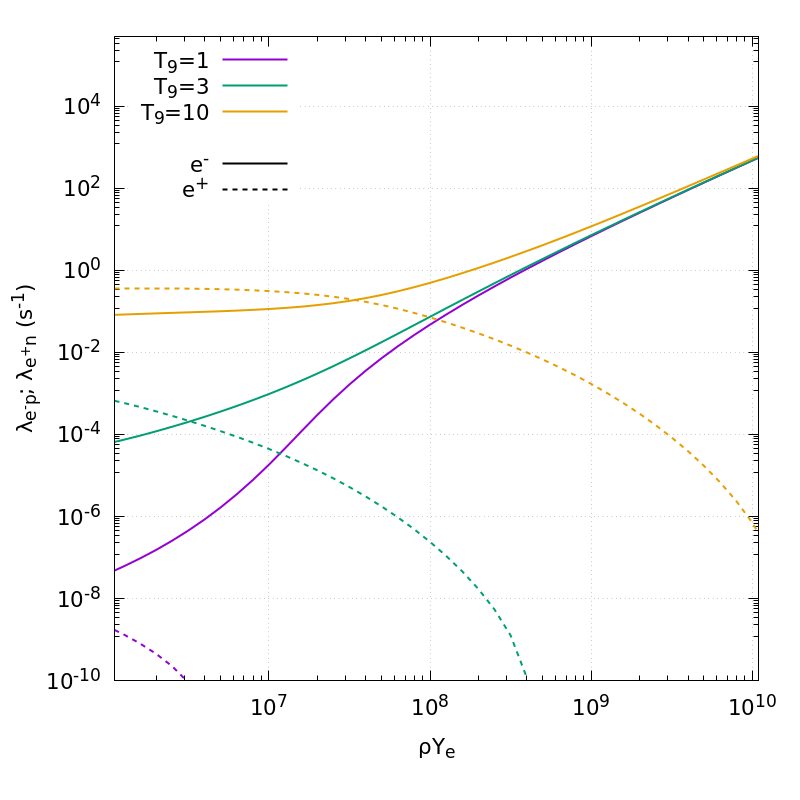}}
\hfil
\caption{Comparison between capture rates $\lambda_{e^{-}p}$ and $\lambda_{e^{+}p}$ at temperatures $T_9=1$, $3$, $10$, and different values of $\rho Y_e$. We note that, out of the low-$\rho$ high-T region, $\lambda_{e^{+}n}<< \lambda_{e^{-}p}$.}
\label{fig:figratespositron}
\end{figure}

\begin{figure}
\centerline{\includegraphics[width=0.5\textwidth]{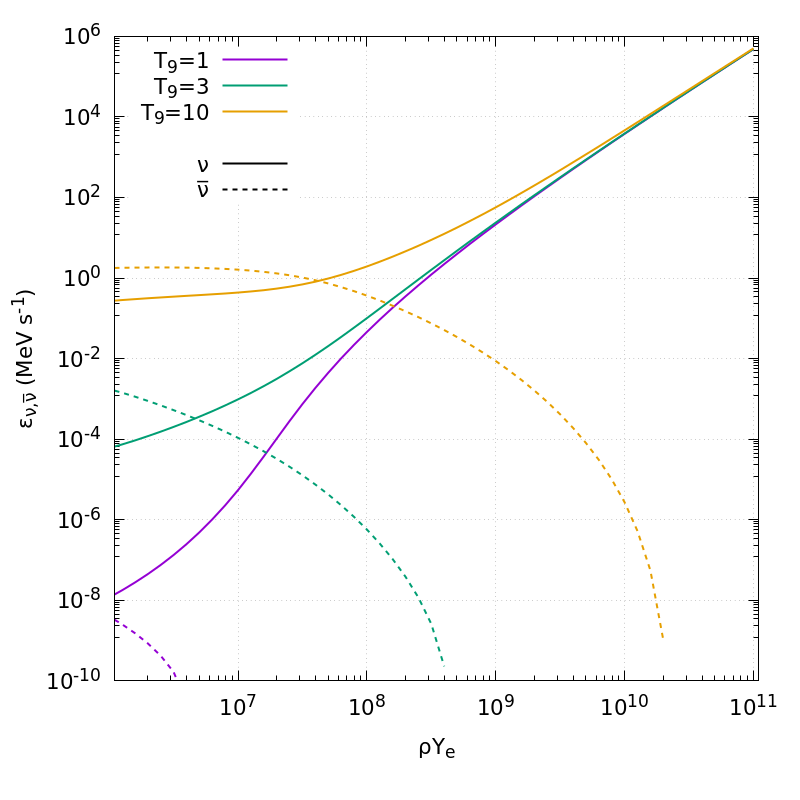}}
\hfil
\caption{Neutrino and anti-neutrino emissivities (in MeV/s) at temperatures $T_9=1$, 3, and 10.}
\label{fig:figneutrinoemission}
\end{figure}

\begin{figure*}
\centerline{\includegraphics[width=\textwidth]{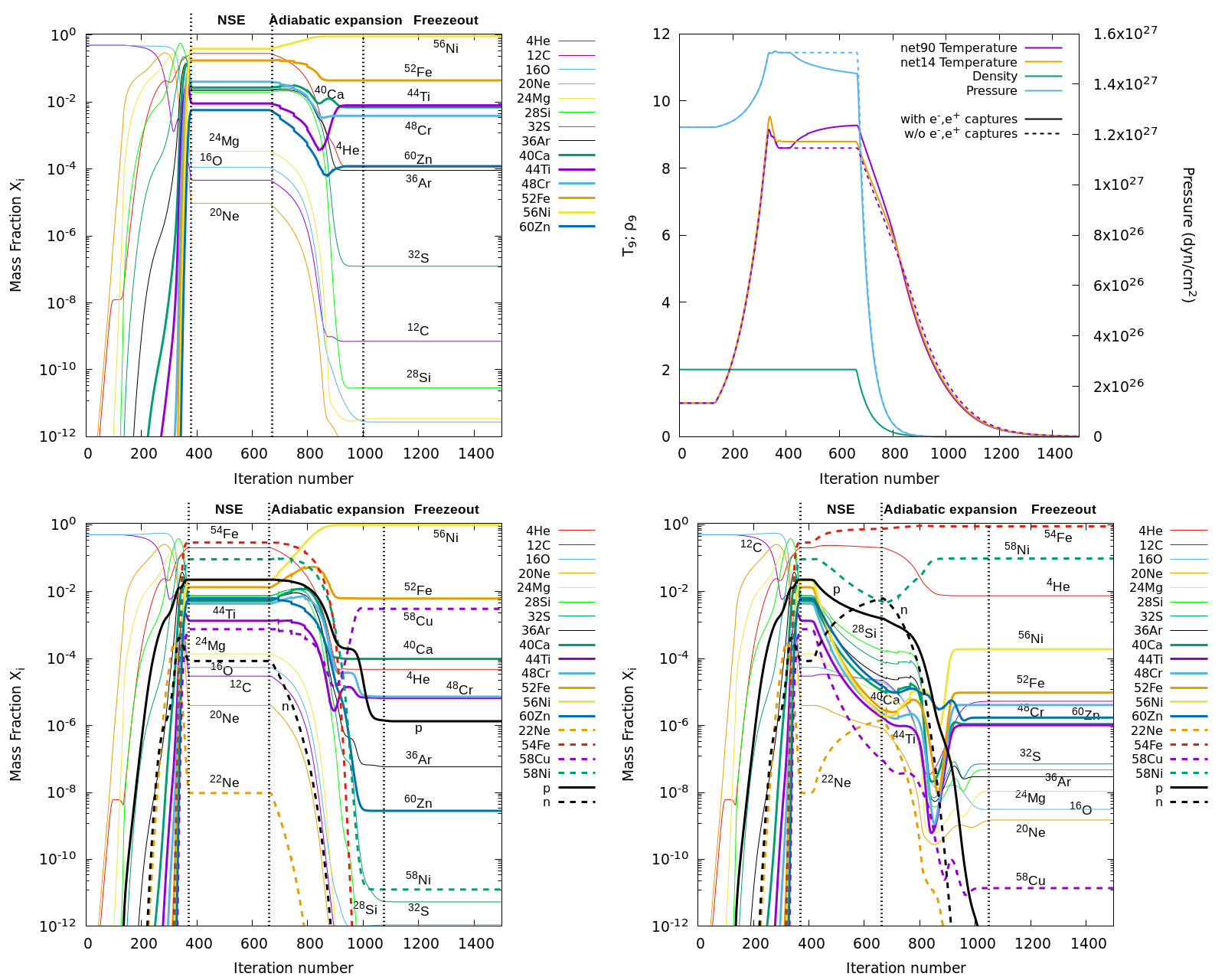}}
\hfil
\caption{Results of the CO test. Evolution of nuclear mass fractions obtained with \netalpha{} (top left), \netnoec{} (bottom left, without \ele{} captures), and \net{} (bottom right, with \ele{} captures on protons included). The vertical dotted lines show the approximate limits of the NSE, when the artificial adiabatic expansion starts, and the region where nuclear reactions are quenched. The panel in the upper right shows the evolution of temperature and pressure for \netnoec{} and \net{}. The temperature obtained with \netalpha{} is also shown as a reference. The evolution of density is the same for all three nuclear networks.}
\label{fig:COtest}
\end{figure*}

\begin{figure}
\centerline{\includegraphics[width=0.5\textwidth]{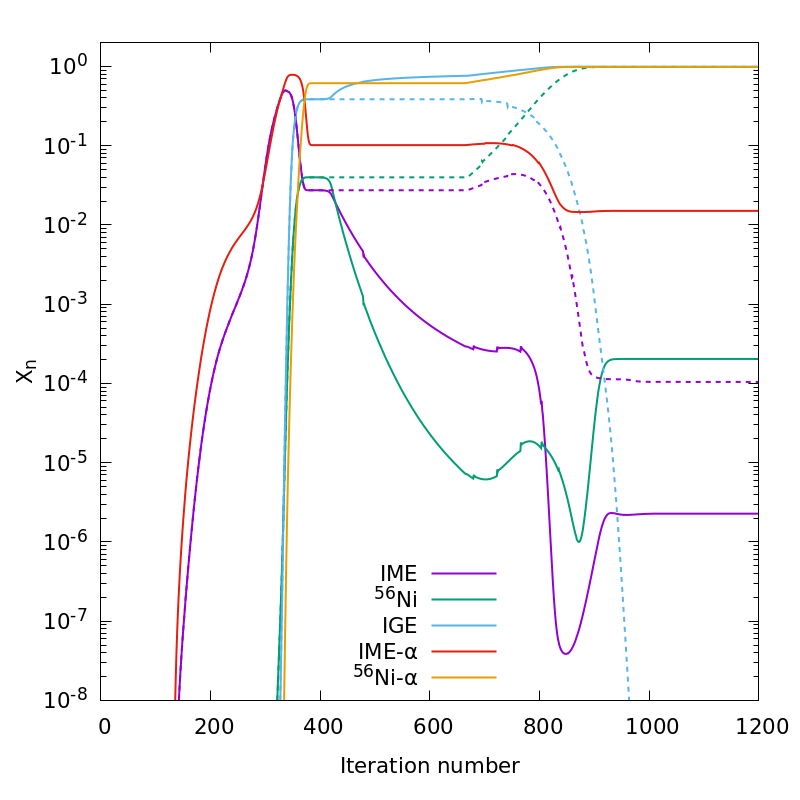}}
\hfil
\caption{Results of the CO test. Mass fractions in Fig.~\ref{fig:COtest}, but grouped in three nuclear families: Intermediate-Mass $\alpha$-Elements (IME, $^{28}$Si$+^{32}$S$+^{36}$Ar$+^{40}$Ca$+^{44}$Ti), $\alpha$-Iron nuclei ($^{48}$Cr$+^{52}$Fe$+^{56}$Ni$+^{60}$Zn, labelled as $^{56}$Ni), and neutronized Iron-Group Elements (IGE, $^{54}$Fe$+^{58}$Ni$+^{57}$Co$+^{58}$Co. Solid lines are for \net{} and dashed lines are for \netnoec{}. We also show two lines calculated with \netalpha{} ($\alpha$-network).}
\label{fig:COGroups}
\end{figure}

\subsection{\net{}}
\net{} is an extension of \neteightsix{} to include electron and positron captures plus three new nuclei and four additional reactions: $^{12}$C($p, \gamma)^{13}$N, $^{13}$N($\alpha,p)^{16}$O, $^{54}$Fe($\alpha, p)^{57}$Co, and $^{57}$Co($n,\gamma)^{58}$Co. The first two have fast reaction rates and become relevant during the explosive combustion of He and the other two work as sinks for $\alpha$ particles and neutrons. This makes \net{} a suitable option for handling moderately asymmetric matter in the context of multidimensional simulations. A sketch of the network is shown in Fig.~\ref{fig:arquitecture_net90}. In the following, we will specify '\emph{net90-noec}' when referring to '\net{} with  $e^{\pm}$ captures turned off', which in turn is very similar to \neteightsix{}.  

In scenarios involving objects such as white dwarfs, where the major contribution to the fluid pressure comes from degenerate electrons, including the electrons in the reduced network is extremely important, and it is something that, as far as we know, has never been done before in the framework of multidimensional hydrodynamic simulations. Indeed, having a process that removes electrons from the fluid, such as capture by protons, can have a significant impact on 3D simulations of Type Ia supernovae. Many reactions involve electrons, but we chose to include only capture by protons, as it is expected to be the dominant process in SNIa \citep{Thielemann1986} and we can benefit from the structure of \neteightsix{} without major changes. Furthermore, and according to \cite{Brachwitz2000} and \cite{Thielemann2004} the \ele{} captures on nuclei are clearly subdominant at densities $\rho_9\simeq 2$, especially since the weak rates compilation of \cite{Langanke2001} significantly reduced the \ele{} capture rates on nuclei of previous compilations \citep{Fuller1982}. We showed and confirmed in the present work (see Sect.\ref{sec:validating}) that the \ele{} captures on free protons alone could amount up to $\simeq 90\%$ of the captures provided that the nuclear network is large enough (f.e. \netWN).

The reaction rate $e^{-}+ p \rightarrow n + \nu$, in a degenerate fluid, can be approached analytically \citep{Hansen1968, Bludman1982} or numerically \citep{fuller1980}. The main advantage of an analytical expression is that its evaluation and its derivative are straightforward, whereas the numerical calculation, although more accurate, requires interpolating from a large data table. We have compared both methods and finally decided to rely on the numerical approach. We show in Figure~\ref{fig:figratescomp} the $e^-$ capture rate, $\lambda_{e^{-}p}$, as a function of $(\rho Y_e)$ for three fiducial temperatures: $T_9=1$, 3, and 10. As we can see, below $\rho Y_e\simeq 10^9$~\dens{} the analytical expressions progressively diverge from the exact numerical value. It should be noted that the expression by \cite{Hansen1968} works well at low density and high temperature, (see f.e. the region $\rho Y_e\le 10^7, T_9=10$), the region where $e^{\pm}$ pair production is relevant.

At temperatures $T_9\simeq 10$ and sufficiently low densities, the rate $\lambda_{e^{+}n}$ of the reaction $e^{+} + n\rightarrow p +\bar\nu$ competes with $\lambda_{e^{-}p}$. We show this in Fig.~\ref{fig:figratespositron} which was obtained by numerically integrating the equations of \cite{fuller1980}.  Although we expect that in a typical SNIa explosion sequence $\lambda_{e^{+}n}< \lambda_{e^{-}p}$, due to the reduction of $T^{max}$ at lower densities (for example, $T_9^{max}\simeq 6, 10$  at $\rho \simeq  2\times 10^7$ and $10^9$\dens{}, respectively), the contribution of $\lambda_{e^{+}n}$ could be significant around $T^{max}$ and we therefore included it in the network.

We assume that neutrinos and antineutrinos created during reactions $(e^{-},p)$ and $(e^{+},n)$ freely escape the system, without further interactions. We computed the $\{\nu, \bar\nu\}$ emissivities with the scheme developed by \cite{fuller1980} and stored the values in an interpolating table. Figure~\ref{fig:figneutrinoemission} shows the expected $\{\nu, \bar\nu\}$ luminosity in the range $10^6$\dens{} $\le \rho\le 10^{10}$\dens{}.

\renewcommand{\arraystretch}{2}
\begin{table*}[ht!]
        \centering
        \caption{Initial conditions for all tests performed with the nuclear networks presented in this work.}
        \begin{tabular}{clccccc}
                \hline
                Test name & \multicolumn{1}{c}{$X_0$} & $T_9$ & $\rho_9$ & $\eta_1$&$\eta_2$&Comment \\
                \hline
                \hline
                \multirow{2}{*}{CO}   & $\left[^{12}\textrm{C}\right]=0.5$  & \multirow{2}{*}{1} & \multirow{2}{*}{2} & \multirow{2}{*}{50}&\multirow{2}{*}{2}& \multirow{2}{*}{WD burning of initially symmetric matter}\\
                     & $\left[^{16}\textrm{O}\right]=0.5$  &  &  & \\
                \hline
                He   & $\left[^{4}\textrm{He}\right]=1$   & 1 & $5\times 10^{-4}$ & {2}& {1}&Shell burning on the WD surface\\
                \hline
                Si   & $\left[^{28}\textrm{Si}\right]=1$   & 6 & 0.01 &{2}& {1}& Photodisintegration leading to CCSN\\
                \hline
        \end{tabular}
        \tablefoot{The columns show (from left to right) the name of the test, the initial composition, the initial temperature and density (in $10^9$ units), and a short comment about the scenario to which the test is relevant.}
        \label{tab:testsICs}
\end{table*}
\renewcommand{\arraystretch}{1}

\begin{figure*}
\sidecaption
\includegraphics[width=12cm]{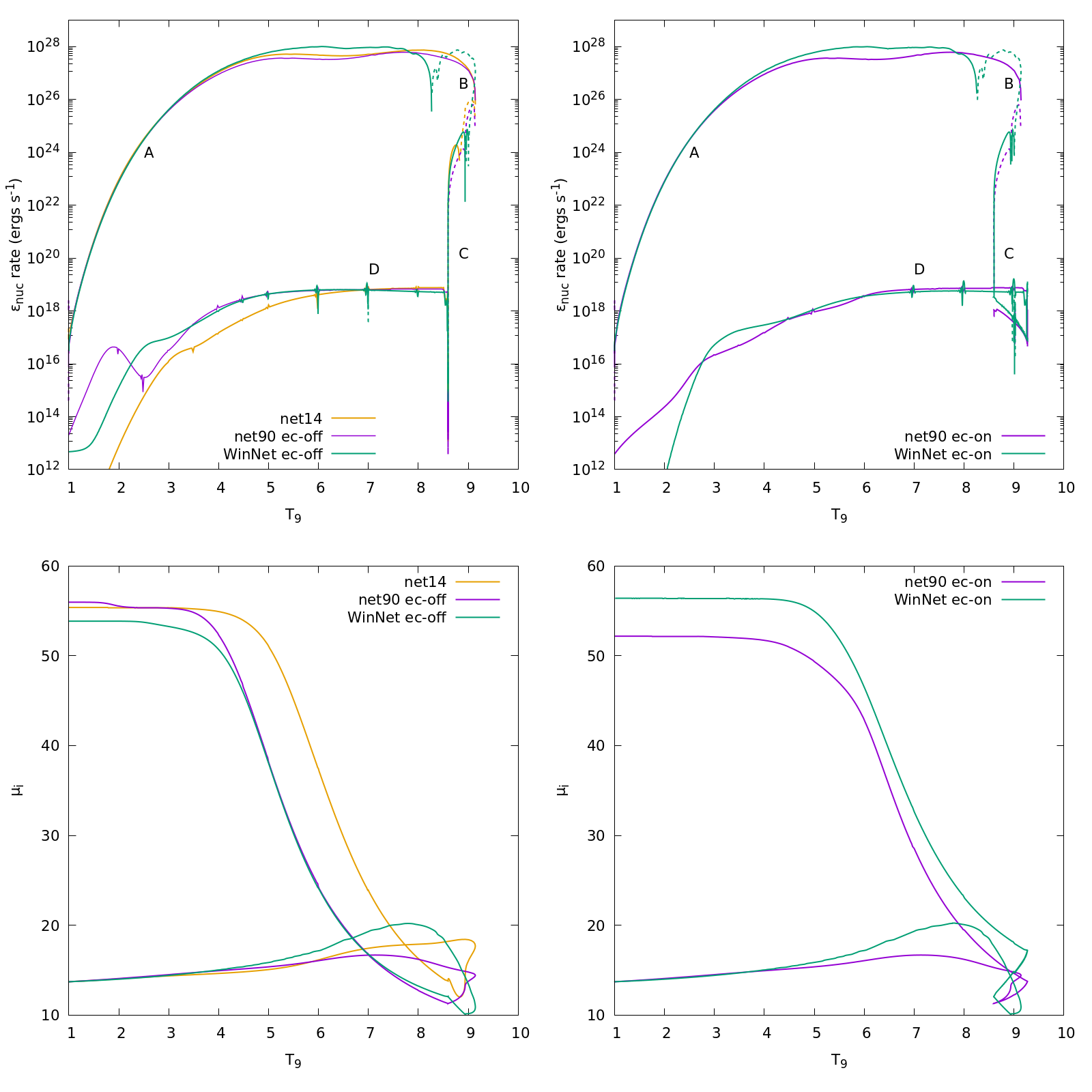}
\hfil
\caption{Results of the CO test. {\sl Top left}: Evolution of the nuclear energy rate as a function of the temperature obtained with \netalpha{}, \netnoec{}, and \netWNnoec{}. {\sl Top right}: Same, but including $e^{-}$ captures. Letters A, B, C, and D refer to normal combustion, relaxation to NSE, full NSE, and adiabatic expansion, respectively. {\sl Bottom left}: Evolution of the mean molecular weight $\mu_i$ of ions as a function of temperature obtained with \netalpha{}, \netnoec{} and \netWNnoec{}. {\sl Bottom right}: Same, but including $e^{-}$ captures.}
\label{fig:COenermuitest}
\end{figure*}

\begin{figure*}
\sidecaption
\includegraphics[width=12cm]{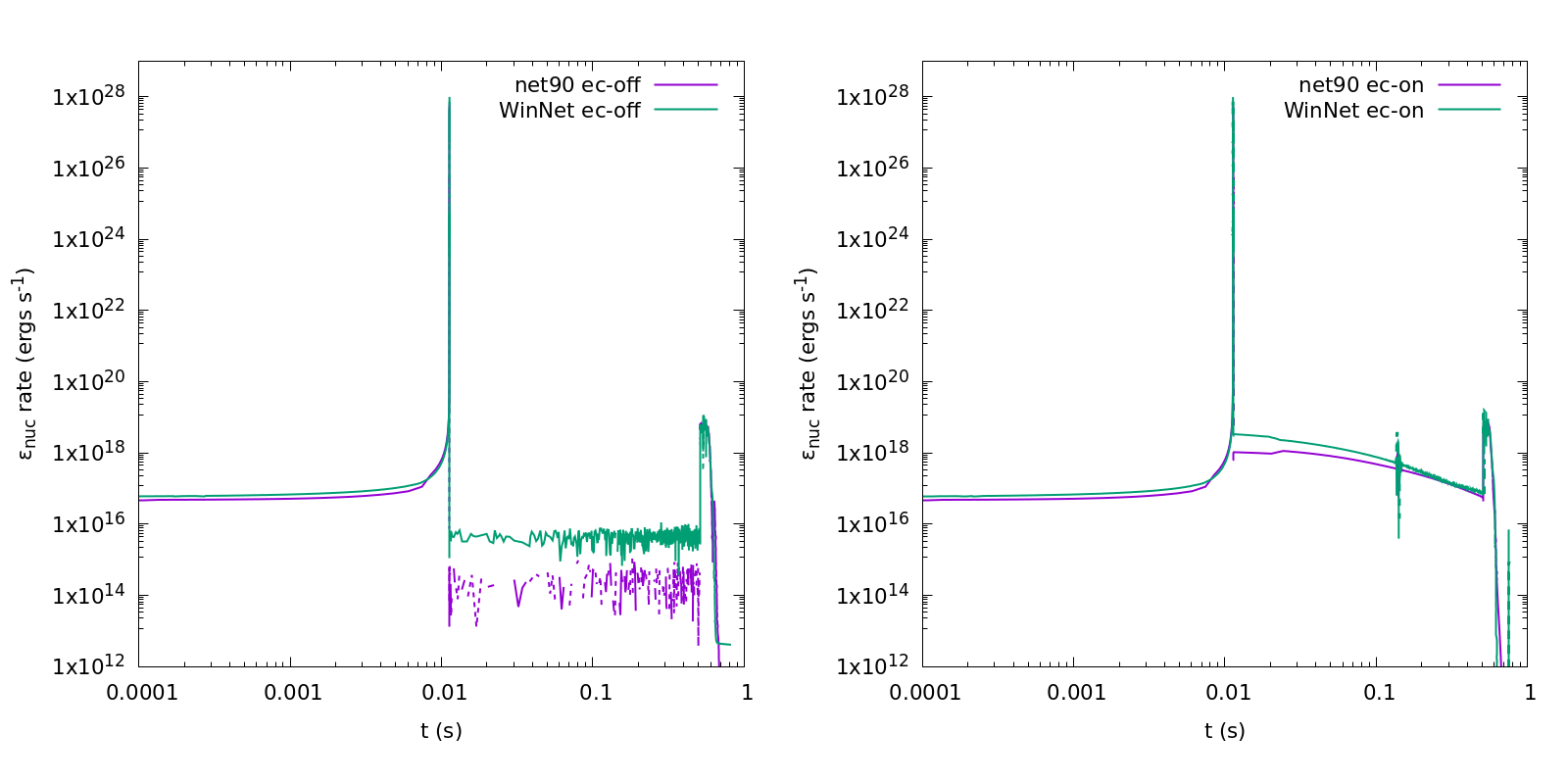}
\hfil
\caption{Time-evolution of the nuclear generation rate in the CO test, computed without (left) and with (right) electron captures.}
\label{fig:testCOnuctime}
\end{figure*}

\subsection{\netWN{}}
The nuclear reaction network \netWN{} is presented in \citet{reichert2023}. Unlike the networks presented previously in this work, \netWN{} contains a flexible number of nuclei. For all calculations, we include 2,281 nuclei up to $Z=50$. This includes all relevant nuclei for the test cases considered. We assume nuclear statistical equilibrium for temperatures higher than 10~GK. Reaction rates have been taken from the REACLIB database \citep{Cyburt2010}. We apply electron screening corrections for all reactions as described in \citet{Kravchuk2014}. Electron and positron capture rates, as well as $\beta^+$- and $\beta^-$-decays that are contained in the REACLIB database, are replaced by those of \citet[][]{Fuller1985,Oda1994,Langanke2001,Pruet2003,Suzuki2016}. Notably, this also includes the electron- and positron-capture rates on free protons and neutrons that are taken from \citet{Langanke2001}. In total, \netWN{} contains $\sim$30,000 reaction rates, of which $\sim$4,000 reactions are weak reactions. Information on the energy of emitted neutrinos was derived from the same tabulation as the theoretical weak rates \citep[][]{Fuller1985,Oda1994,Langanke2001,Pruet2003,Suzuki2016} and from the ENSDF database \citep{Brown2018}, when it was unavailable in previous tables. Also, we consider the energy of thermal neutrinos \citep{Itoh1996}. 

\section{Tests}
\label{sec:tests}
In this section, we present a suite of tests in which, given an initial composition, temperature, and density, we iteratively apply our nuclear networks with a self-adaptive timestep until NSE or a stable configuration has been achieved isochorically. This usually happens faster than the hydrodynamic timescale, defined as:

\begin{equation}
t_{hd}=446 / \sqrt{\rho_0},
\label{Eq:thd}
\end{equation}

\noindent
where $\rho_0$ is the initial density value (corresponding to $\rho_9$ in Table~\ref{tab:testsICs}). Once the physical time is larger than $t = \eta_1 t_{hd}$, where $\eta_1$ is a constant that depends on the particular scenario, we impose an artificial exponential decrease in density, mimicking an adiabatic expansion of the fluid element in which the nuclear reactions are taking place,

\begin{equation}
\rho(t)=\rho_0\exp{\left(-\frac{\Delta t}{\eta_2 t_{hd}}\right)},
\label{Eq:rhoexp}
\end{equation}

\noindent
where $\Delta t$ is the current timestep and the choice of $\eta_2$  depends on the particular calculation. This produces a cooling that naturally leads to the freezeout of the nuclear reactions. In Table~\ref{tab:testsICs} we summarize the initial conditions of all the tests presented in this section. We used the EOS of \cite{Timmes2000b} to monitor the temperature changes and pressure in all tests.

\begin{figure*}
\sidecaption
\includegraphics[width=12cm]{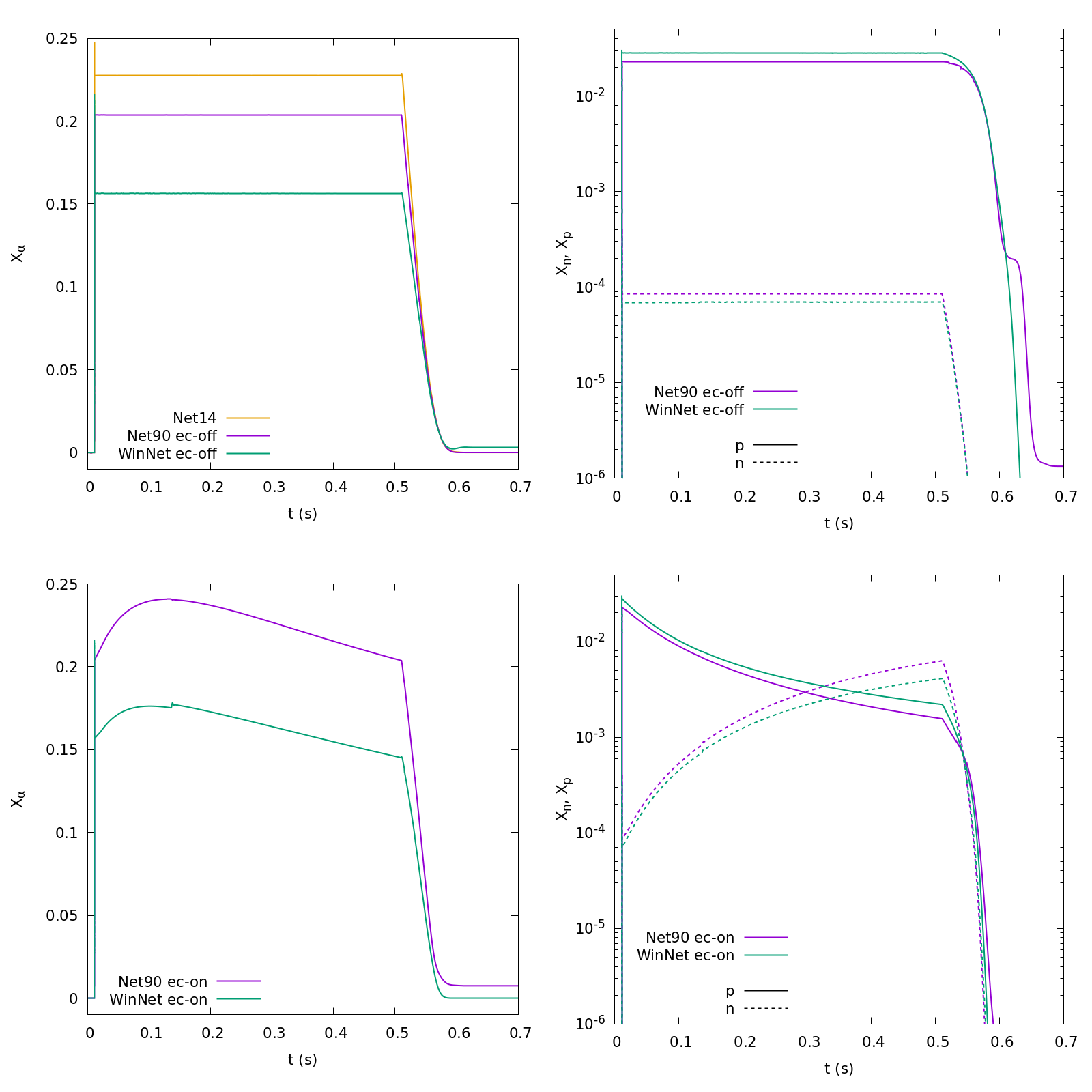}
\hfil
\caption{Results of the CO test. {\sl Top left}: Evolution of the mass fraction $X_\alpha$ of alpha particles of \netalpha{}, \netnoec{}, and \netWNnoec{}. {\sl Top right}: Same, but for neutron and proton mass fractions. {\sl Bottom left}: Evolution of the mass fraction $X_\alpha$ of the alpha particles of \net{} and \emph{WinNet}. {\sl Bottom right}: Same, but for proton and neutron mass fractions.}
\label{fig:COXpnatest}
\end{figure*}

\subsection{C+O burning}
This test is characteristic of a Chandrasekhar mass white dwarf core that undergoes a nuclear runaway that leads to a type Ia supernova explosion. We used an equal amount of \carbon{} and \oxygen{} to mimic the gross conditions prior to the explosion. The initial conditions for this test are summarized in Table~\ref{tab:testsICs}. Under these conditions, NSE is achieved rapidly when the temperature increases and the radiative captures of protons, neutrons, and alpha particles are balanced by photodisintegrations. The chosen value of $\eta_1$ in Table~\ref{tab:testsICs} is compatible with the duration of the NSE combustion regime in the center of a massive white dwarf that results in explosive ignition of carbon. 

Figure~\ref{fig:COtest} presents the results of the calculation. The top-left panel shows the evolution of the alpha-chain elements obtained with \emph{net14}. NSE is achieved around iteration 380. The adiabatic expansion sets in around iteration 660 until the density and temperature are so low that all nuclear reactions freeze out around iteration 1000, with a final composition dominated by $^{56}$Ni (followed by $^{52}$Fe) as it is the dominant nuclei in Type Ia supernova explosions.

The bottom left panel of Fig.~\ref{fig:COtest} shows the nuclear evolution when we use \netnoec{}. To ease comparison, the colors of all alpha-chain elements are the same in all plots, except that in the larger networks we also show a few (most abundant) isotopes in colored, dashed lines, along with proton and neutron abundances (in black). For most of the pre-NSE regime, nuclear mass fractions follow an evolution similar to that obtained by \netalpha{}. Only when the temperature is high enough ($T_9\ge 7$), many new production channels open, given the high abundance of fresh protons and neutrons. This leads to a longer quasi-NSE regime (compared to \netalpha{}) and a completely different distribution of abundances in NSE, dominated by moderately neutronized isotopes such as $^{58}$Ni and $^{54}$Fe, as well as protons and neutrons. After the adiabatic expansion sets in, most of $^{58}$Ni captures a proton to become $^{59}$Cu, which produces $^{56}$Ni through an inverse $(\alpha,p)$ reaction. Another path for $^{58}$Ni is two consecutive inverse $(n,\gamma)$ reactions, but this is subdominant due to the rapid cooling during the expansion and evidenced in the very fast decrease in neutron abundance. At freezeout, the most abundant element is again $^{56}$Ni, followed by some of the heavier isotopes in the network.
The final mass fractions of $\alpha-$particles calculated with \netalpha{} and \netnoec{} are quite similar and low, $X_\alpha\simeq 10^{-4}$, slightly larger in the former.

The bottom right panel of Fig.~\ref{fig:COtest} corresponds to the results obtained by \net{}. That is, including electron captures. As expected, there are no large differences until the temperature rises enough to trigger the electron captures on protons. This has a major impact on the NSE, where the proton abundance drops by more than an order of magnitude, producing a large abundance of neutrons. Reactions with neutrons progressively reduce the mass fraction of $\alpha-$elements favoring the synthesis of moderately neutron-rich elements. Now the NSE is largely dominated by $^{54}$Fe (shown as $^{54}$Fe+$^{57}$Co+$^{58}$Co in Fig.~\ref{fig:COtest}) followed by $^{58}$Ni. Unlike nuclei, electron captures never achieve equilibrium at the densities considered in this work, and neither does the temperature. The larger binding energy of the dominant neutron-rich isotopes increases the released nuclear energy so that the temperature actually increases during the NSE stage. 

This is clearly shown in the top right panel of Fig.~\ref{fig:COtest}. There, we show the evolution of temperature ($T_9$), density ($\rho_9$), and pressure. The dashed lines correspond to \netnoec{}, where no electron captures were included. The solid lines correspond to \net{}, including electron captures, with the exception of the orange solid line, which represents the temperature evolution obtained with \netalpha{}, as a reference. Comparing lines with the same color shows the influence of including electron captures in the nuclear network. This leads to a relevant increase in temperature (at NSE and a good fraction of the adiabatic expansion), whereas pressure gets lower as electrons are captured by protons. In this test, $Y_e$ decreased from 0.5 to 0.472. As discussed in Sect.~\ref{impactCh}, this effect may have an appreciable influence on the dynamical evolution of hot and dense systems in multidimensional simulations where nuclear reactions are relevant. 

{Figure~\ref{fig:COGroups} shows the mass fractions of IME, IGE, and $^{56}$Ni-like elements (see its caption for a definition of these groups). The inclusion of \ele{} captures produces substantial changes in the abundances after iteration 400. In particular, the production of IME and $^{56}$Ni is severely overestimated when \ele{} captures are not taken into account. The opposite is true for the IGE elements, which are under-produced, especially during the adiabatic expansion. The reduced \netalpha{} produces even more IME and $^{56}$Ni than \netnoec{}. Without \ele{} captures the abundance at freezing is dominated by $^{56}$Ni whereas the simple inclusion of \ele{} captures on protons shifts the dominance to neutronized IGE elements.} 

\subsubsection{Verifying \net{} with \netWN{}}
\label{sec:validating}

We present here a comparison between \net{} and the reference nuclear network \netWN{} in light of the magnitudes with the largest impact on the hydrodynamics, such as the rate of released nuclear energy $\dot\epsilon_n (t)$, pressure $P(t)$ and values of $Y_e (t)$ and mean molecular weight of ions $\mu_i (t)$. To facilitate comparisons among \netalpha{}, \netnoec{} and the larger network but without \ele{} captures, \netWNnoec{}, we first calculate the evolution with \netnoec{} and use the resulting trajectory $\{T(t), \rho(t)\}$ as input to the \emph{net14} and \netWNnoec{} calculations. The same procedure is used to compare \net{} and \netWN{}, now with \ele{} captures. We stress that although \net{} provides a general picture of the abundances of many isotopes during the run time of multidimensional hydrodynamic calculations, the post-processing of the data is still necessary to obtain detailed nucleosynthetic yields.    

Figure~\ref{fig:COenermuitest} shows the value of $\dot\epsilon_n$ and the mean molecular weight of ions $\mu_i$ as a function of temperature for the different nuclear networks at ignition density $\rho=2\times 10^9$ g$\cdot$cm$^{-3}$ and initial composition $X(^{12}\textrm{C})=X(^{16}\textrm{O})=0.5$.  The evolution of $\dot\epsilon_n$ calculated without and with $e^{-}$ captures during the initial rapid temperature increase (labeled with letter A in the uppermost panels) is fairly similar. In this region and below $T_9=4$ the three networks produce identical results, whereas \netWNnoec{} releases a bit more energy between $4\le T_9\le 8$. This is followed by a rapid relaxation to the NSE regime (region B), where endothermic photodisintegrations take over (dashed lines in the figures).  During nuclear statistical equilibrium (region C), the temperature remains close to $T_9=8.6$ and the nuclear energy generation rate drops several orders of magnitude. In this stage, calculations with $e^{-}$ captures release more nuclear energy than those that do not consider the effect of $e^{-}$ captures (see also Fig.~\ref{fig:testCOnuctime}). The match between \net{} and \netWN{} is quite good in this region, and \net{} is able to reproduce the loop around $\dot\epsilon_n=10^{18}$ ergs$\cdot$s$^{-1}$ in the upper right panel of Fig.~\ref{fig:COenermuitest}. The agreement also holds for a large part of the adiabatic expansion (zone D), where both networks give comparable results between $4\le T_9\le 8.5$. Note that below $T_9\simeq 6$ the reduced $\alpha$-network \netalpha{} differs considerably from the results of \netnoec{} and \netWNnoec{} (top left panel in Fig.~\ref{fig:COenermuitest}). 

The impact of including \ele{} captures in the network is evident in Fig.~\ref{fig:testCOnuctime} which shows the time evolution of the nuclear energy generation rate. Before NSE, at $t\le 0.01$~s, the rate is practically the same in all cases, but the behavior is different thereafter, with calculations including \ele{} captures providing a considerably larger (and positive) generation rate. Furthermore, \netnoec{} produces practically no energy during the equilibrium phase when $t>0.01$~s, showing very low and alternate $\pm$ values, as expected. On the other hand, \netWNnoec{} releases more energy with a positive sign (left panel), while during the NSE \netWN{} initially releases more energy, but the difference with \net{} gradually decreases, becoming negligible when $t>0.1$~s (right panel). 

The mean molecular weight of ions $\mu_i$, shown in Fig.~\ref{fig:COenermuitest}, has a maximum relative error $\simeq 20\%$ between \net{} and \emph{WinNet} during the fast relaxation to NSE at $T_9\simeq 9.3$. However, such a relative difference is much lower in other combustion stages. During freeze-out, the relative difference decreases to $\simeq 4\%$ and $\simeq 8\%$ in the calculations without and with electron captures, respectively. During the adiabatic expansion, there is almost a perfect match between \netnoec{} and \netWNnoec{} while \netalpha{} differs significantly (bottom left panel). According to the bottom right panel of Fig.~\ref{fig:COenermuitest} the value $\mu_i$ calculated with \net{} remains close but below that of \netWN{}, which is attributable to the different mass fractions of helium obtained with the two networks, as commented below. 

Figure~\ref{fig:COXpnatest} depicts the evolution of light elements, $n, p$, and $\alpha$ particles. The latter (leftmost panels) are the most abundant during NSE in all cases, with \netnoec{} giving values closer to \netWNnoec{} than \netalpha{}. Models implementing \ele{} captures produce more helium during the NSE due to the higher temperature and leave more uncombined $\alpha$ particles at freezing (magenta lines in the leftmost panels in Fig.~\ref{fig:COXpnatest}) The evolution of neutrons and protons (rightmost panels) is fairly well reproduced by \net{} in all cases. 

\begin{figure}
\centerline{\includegraphics[width=0.5\textwidth]{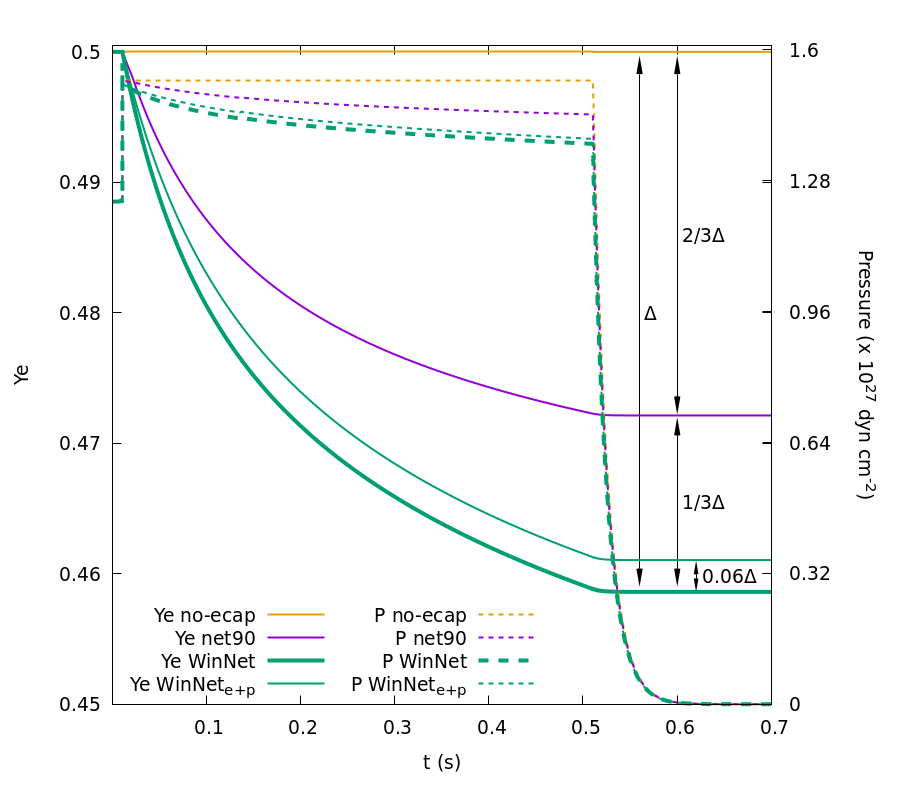}}
\hfil
\caption{Results of the CO test. Evolution of the electron molar fraction and pressure calculated with \netnoec{}, \net{}, \emph{WinNet}, and \netWN{}$_{{\mathrm e+p}}$ with electron captures only on nucleons (i.e. only with $e^-+p$ and $e^++n$ reactions and no electron captures on heavier nuclei).}
\label{fig:COYePresstest}
\end{figure}

Figure~\ref{fig:COYePresstest} shows the evolution of the molar fraction of electrons, $Y_e$, which has a major impact on pressure, which in turn is a crucial magnitude in hydrodynamic simulations. The evolution of pressure is shown on the right-hand Y-axis of the same figure. As we can see, including \ele{} captures, if only on free protons, is worthwhile, as it accounts for $\simeq 2/3$ of the total amount of captures. Figure ~\ref{fig:COYePresstest} also shows that the contribution of \ele{} captures on nuclei is actually quite small $\simeq 6\%$ (distance between the thick and thin green solid lines). This suggests that both, the size and the architecture of the reduced networks, are relevant to depict the evolution of $Y_e$ and that the thin green line calculated with \netWN{} but excluding \ele{} captures on nuclei, represents a limiting case. As shown in the same figure, the strong dependence of the electron pressure on $Y_e$, $P\propto Y_e^{4/3}$, makes \ele{} captures relevant to hydrodynamic simulations of Chandrasekhar-mass models of SNIa. 

\begin{figure*}
\sidecaption
\includegraphics[width=12cm]{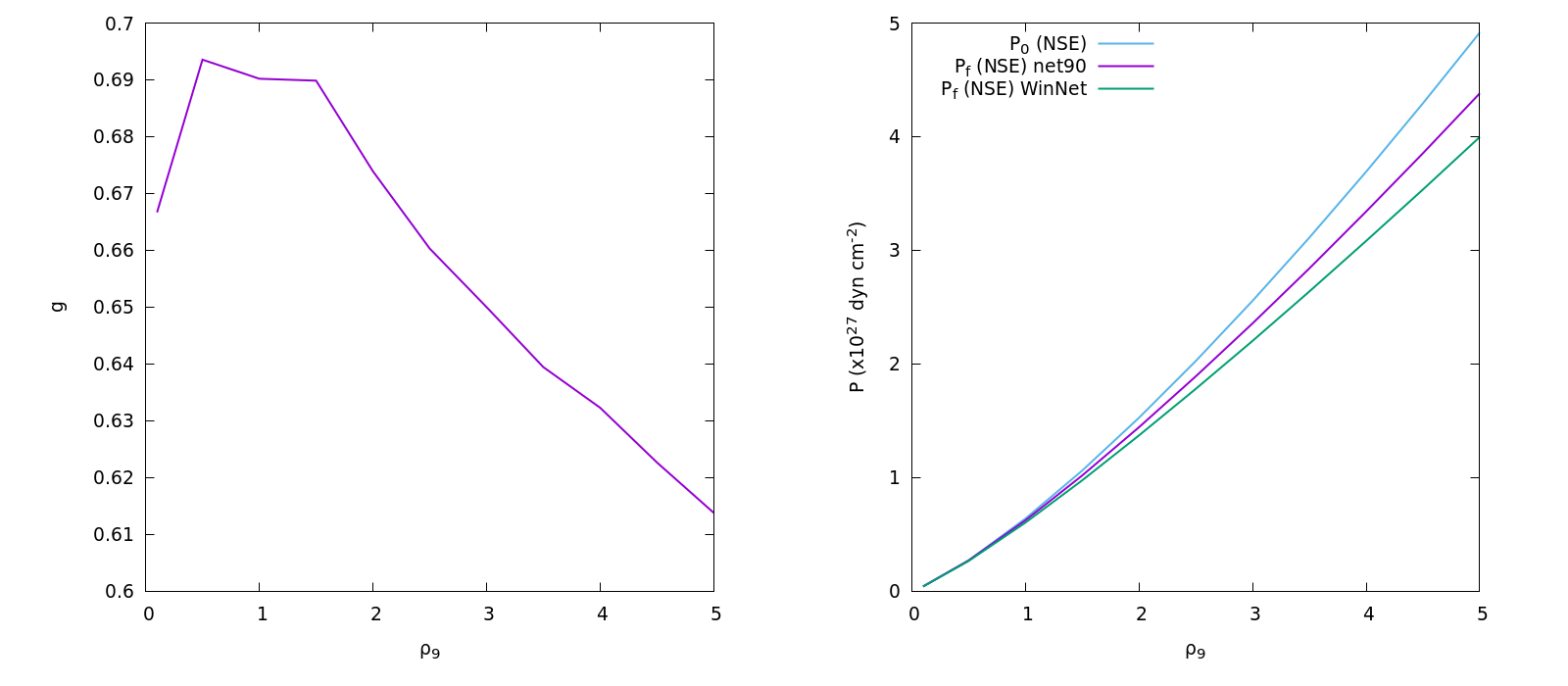}
\hfil
\caption{Results of the CO test. Left: profile of $g=[\Delta Y_e (\net{})/\Delta Y_e (\netWN{})]$ at the freeze-out time as a function of the ignition density. Right: profile of total pressure at the beginning of NSE, $P_0$, and at the end point of NSE, $P_f$, when the adiabatic expansion begins. }
\label{fig:COyepresscomp}
\end{figure*}

The impact of changing the ignition density is explored in Fig.~\ref{fig:COyepresscomp}. At each density and in the range $1\times 10^8 \le \rho \le 5\times 10^9$\dens{} the burning temperature is first estimated with \net{} assuming isochoric combustion that lasts the same amount of time as in the reference case with $\rho=2\times 10^9$\dens{} studied above. The system is then allowed to expand with the characteristic time of the reference case in Table~\ref{tab:testsICs}. The resulting $\rho(t), T(t)$ profiles are used as input for \netWN{} to estimate more accurate values of $Y_e$ and pressure. The left panel of Fig.~\ref{fig:COyepresscomp} shows the dependence of

\begin{equation}
g=\frac{[Y_e(0)-Y_e^{\net{}}(t_f)]}{[Y_e(0)-Y_e^{netWN}(t_f)]}\,,
\label{eq:Yedens}
\end{equation}

\noindent as a function of the ignition density, where $t_f$ is the freeze-out time. The estimator $g$ gives the relative fraction of \ele{} captures estimated with \net{} compared to those with \netWN{}. Although the value of $g$ decreases with $\rho_9$, it always remains above $63\%$. The slight reduction of $g$ at high densities is due to the progressive relevance of electron captures on neutronized nuclei in \netWN{}. The impact on pressure is shown in the right panel of Fig.~\ref{fig:COyepresscomp} where the depressurization is evident (i.e. the separation from the blue line), especially at high densities. Despite its great simplicity, \net{} does a good job here because it can account for more than $60\%$ of the pressure reduction predicted by \netWN{}. We note in passing that the final values of $Y_e$ always lie above the theoretical limit obtained by \cite{Arcones2010} assuming complete beta equilibrium.

\subsubsection{Impact on Chandrasekhar-mass explosion models of SNIa}
\label{impactCh}

Although the explosion of a white dwarf with a mass approaching the Chandrasekhar-mass limit was historically the favored progenitor to explain Type Ia supernova explosions, it is today thought not to be the dominant production channel for these events \citep{mao14}. However, such a channel may still account for a few percent of these explosions, which can increase to $20\%$ provided that the ignition density range is extended to $\simeq 6\times 10^9$\dens{} \citep{bravo2022}, while a more standard ignition value is $1-3\times 10^9$\dens{} \citep{nom84}. In any case, this interval of densities is within the working range of \net{}. 

Any reduction in $Y_e$ caused by \ele{} captures has, to a greater or lesser extent, an impact on the dynamics of the explosion. However, the current reduced or medium-sized nuclear networks used to perform multidimensional simulations of SNIa explosions do not include deleptonization reactions \citep[e.g.,][]{Ropke2005}\footnote{Another option is to consider big tables, previously built with a large network that include weak processes, which store the generation of energy and composition changes from which the code makes the necessary interpolations during the run-time of the calculation \citep{moll2013}  }. 

In matter dominated by degenerate electrons and in the ultra-relativistic regime, the electron pressure is dominant and approached by $P_e=k (\rho Y_e)^{4/3}$, thus:

\begin{equation}
    \left(\frac{\delta P_e}{P_e}\right)_{\rho} = \frac{4}{3}\left(\frac{\delta Y_e}{Y_e}\right)\,,
    \label{Eq:deltaP}
\end{equation}

\noindent which for $Y_e=0.5; \delta Y_e = -0.03$ gives $\delta P_e/P_e=- 0.08$ which although not large, contributes to delay the expansion of the fluid element.

Furthermore, deleptonization affects the propagation of thermonuclear combustion. It is commonly accepted that in Chandrasekhar-mass models the thermonuclear combustion is initially propagated by convective motions, where hot bubbles filled with combustion products and formed at low radius float and disseminate the combustion at higher altitudes \citep{niemeyer1997}. The ascending velocity of these bubbles is governed by the expansion factor $\mu=\rho_b/\rho_u$ between burnt and unburnt matter. The deleptonization caused by electron captures accelerates the pressure balance between the hot bubble and the cool surroundings and increases the value of $\mu$, thus reducing the rising velocity of the hot blobs. 

We have checked the efficiency of \net{} in handling these Chandrasekhar-mass models by tracking the thermonuclear combustion of the central layer of a white dwarf that expands with a density profile $\rho(t)$ adopted from the W7 model by \cite{nom84}. To do that, we first calculate the self-consistent NSE state at $\rho(t=0)$ with \net{} and follow the evolution of the temperature and nuclear species as the density declines, following W7, until the complete freeze-out of the mixture. The $T(t)$ profile obtained with \net{} is used to postprocess the output with \netWN{}.

\begin{figure*}
\sidecaption
\includegraphics[width=12cm]{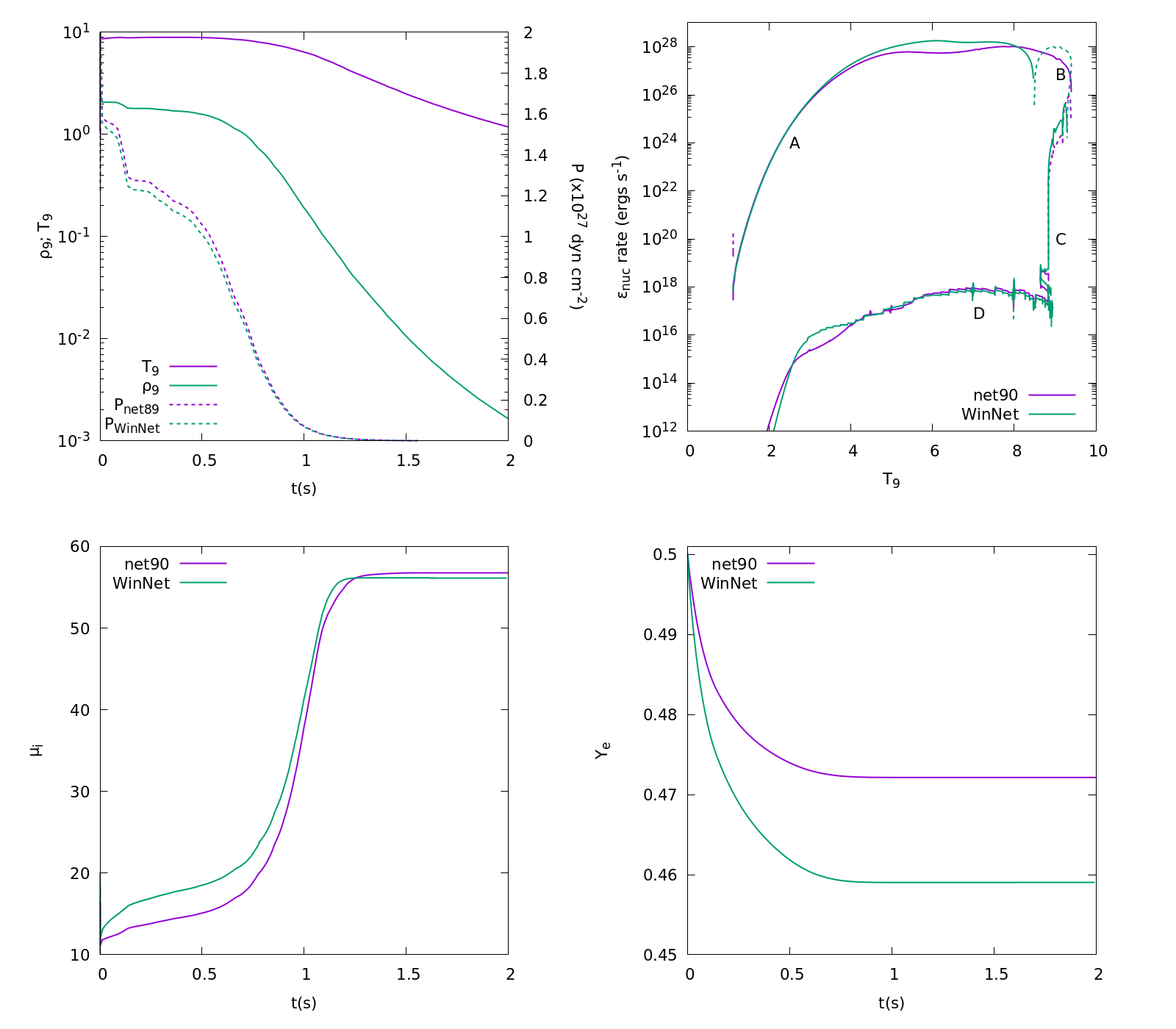}
\hfil
\caption{Results of the W7 test. {\sl Top left}: Evolution of temperature and pressure for the density evolution of the central region of the W7 model, {\sl Top right}: Evolution of the rate of released nuclear energy obtained with \net{} and \netWN{} as a function of $T_9$. Letters A-D refer to normal combustion, relaxation to NSE, full NSE, and adiabatic expansion, respectively. {\sl Bottom-left}: Evolution of the mean molecular weight $\mu_i$ of ions obtained with \net{} and \netWN{} . {\sl Bottom right}: Evolution of the electron molar fraction $Y_e$ $\mu_i$ obtained with \net{} and \netWN{} }
\label{fig:COWW7Test}
\end{figure*}

We provide a summary of the results in Fig.~\ref{fig:COWW7Test} which shows the evolution of density, temperature, and pressure (top-left panel), released nuclear energy rate $\dot\epsilon_n$ (top right), mean molecular weight of ions (bottom left), and electron molar fraction (bottom right), respectively. These results are qualitatively similar to those of Fig.~\ref{fig:COenermuitest} and the same comments given in Sect.~\ref{sec:validating} regarding $\dot\epsilon_n$ apply here. The evolution of $\mu_i$ matches \netWN{} fairly well, especially at $t\ge 0.8$~s and \net{} is capable of accounting for a good amount of \ele{} captures ($\simeq 2/3$ of those with \netWN{}). 

Table~\ref{tab:table_W7} provides quantitative information of several magnitudes at $t=2$ s at which any nuclear activity has ceased. The final abundance of protons and neutrons is negligible in both calculations. There is some mismatch in the mass fraction of $\alpha$ particles, but the final value with \net{} is low enough to still not affect the molecular weight of ions and hence the ionic pressure. This tiny amount of remaining alphas is due to the limited number of $\alpha$ reactions with neutron-rich nuclei included in \net{}. The final value of $\mu_i$ is well reproduced with the reduced network, and the most abundant nuclei is $^{57}$Co with an atomic mass and a mass fraction similar to $^{56}$Fe obtained with WinNet. The total energy released at $t=2$~s is in agreement within $4\%$. Note that according to the last column in Table~\ref{tab:table_W7} the energy released by the freely escaping neutrinos is not negligible and with $E_\nu(\net{})/E_\nu(\netWN{})\simeq 67\%$. This allows us to monitor the approximate neutrino display during the run-time calculation of the explosion.        

\begin{table*}
        \centering
        \caption{Results of the W7 test.} 
       \begin{tabular}{cccccccccc} 
               \hline
                
               Network& $X_p$&$X_n$& $X_{\alpha}$& $X_{max}$& $\mu_i$&$ Y_e$ &  $\epsilon_{n}$&$E_{\nu}$& \\
               &&&&& && $\times 10^{17}$ ergs$\cdot$g$^{-1}$ &$\times 10^{17}$ ergs$\cdot$g$^{-1}$ &  \\
                \hline
                \hline
               \net{} & $0$ & $0$ & $7.2\times 10^{-4}$& 0.56 ($^{57}$Co)  & $56.8$ &$0.472$& $8.57$& $1.2$\\
                \netWN{} & $0$ & $0$  & $10^{-10}$& 0.53 ($^{56}$Fe)& $56.2$&$0.459$& $8.95$& $1.8$\\
                
                \hline
    
       \end{tabular}
       \tablefoot{Evolution of the central layer of the W7 model \citep{nom84}. We show several magnitudes calculated with \net{} and \netWN{} at $t=2$~ s, when complete freeze-out of the reactions has been achieved. Columns are: the name of the nuclear network, the mass fractions of protons, neutrons, $\alpha$ particles, and most abundant nuclei, the mean molecular weight of ions, electron molar fraction, total released nuclear energy, and energy carried out by freely escaping neutrinos, respectively.}
       \label{tab:table_W7}
\end{table*}

\subsubsection{The deflagration of a white dwarf with \net{} and calculated in three-dimensions}
\label{sec:impactCh3D}

We use \net{} to simulate the deflagration of a WD in a true three-dimensional environment. Our goal is to show how the implicit network can handle the explosion of a massive white dwarf made of $50\%$ carbon and oxygen and to evaluate the impact of the \ele{} captures on the outcome of the deflagration. We simulate the explosion of a $1.376$~\msun{} WD ($\rho_c=2.9\times 10^9$\dens{}) following the artificial incineration of six small bubbles in the central volume of the WD, at random distances $r\le 80$ km from the center and aligned with the three coordinate axis. We perform three-dimensional low-resolution calculations with the smoothed particle hydrodynamics (SPH) code SPHYNX \citep{cab17} adapted to handle nuclear flames and deflagrations, and whose features will be reported elsewhere \cite[see also][for a description of the nuclear flame propagation algorithm]{garcia-senz2016}. We conducted two simulations with 1 million of SPH particles, including and neglecting electron captures in \net{}.

\begin{figure*}
\sidecaption
\includegraphics[width=12cm]{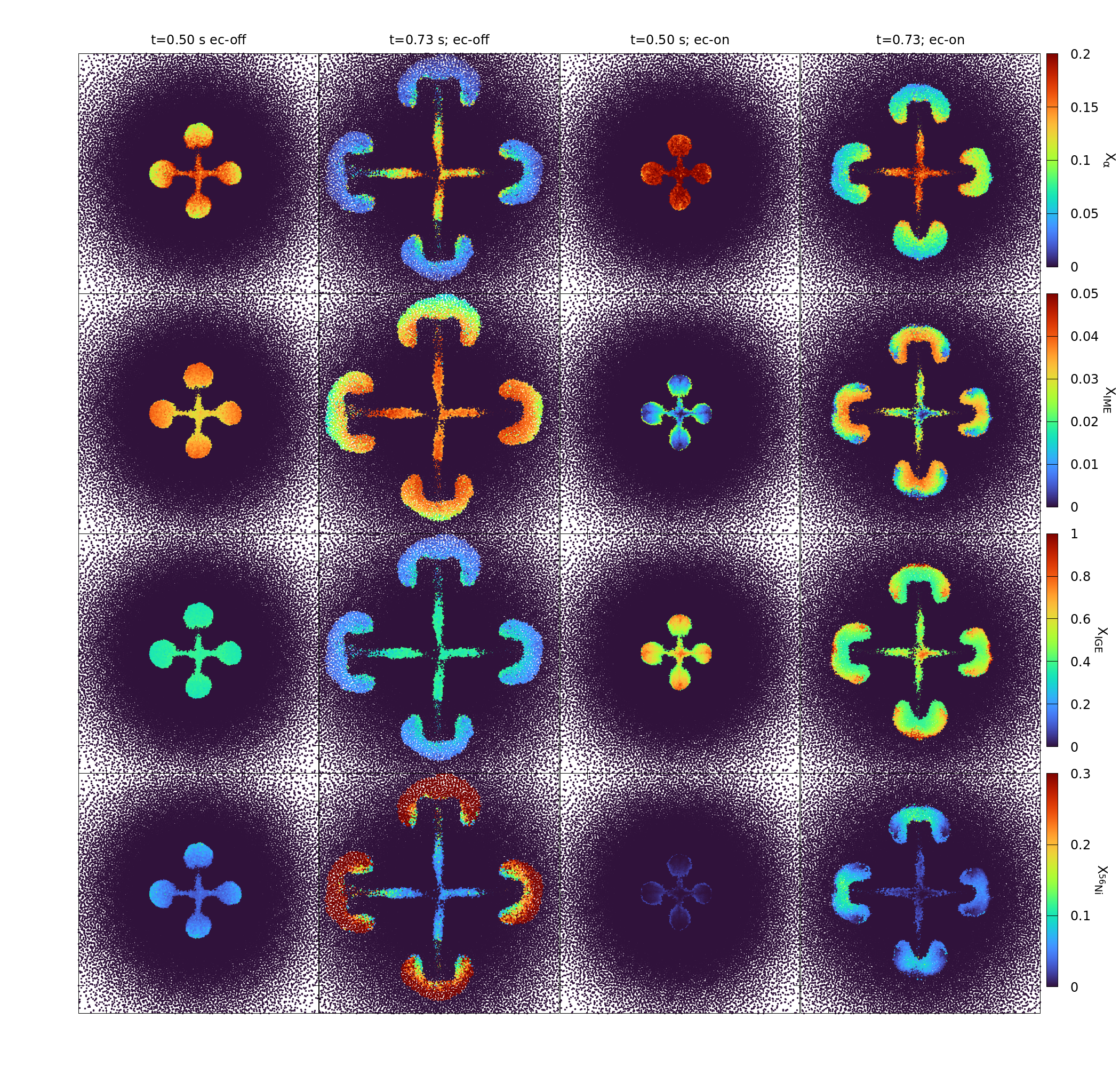}
\hfil
\caption{ XY slice showing the distribution of four groups of isotopes during the deflagration of a WD at two elapsed times: $X_\alpha$ (upper row), intermediate-mass elements, $X_{IME}$ (second row, see Fig.~\ref{fig:COGroups} for a definition of these groups), iron-group elements $X_{IGE}$ (third row), and $^{56}$Ni (last row). The first two columns were obtained neglecting \ele{} captures whereas the other two include them. Each box has a side of $1300\times1300$~km.}
\label{fig:3DTypeSNIanuc}
\end{figure*}

\begin{table*}
        \centering
        \caption{Final yields during the deflagration of a white dwarf calculated in three-dimensions}
       \begin{tabular}{ccccccc} 
               \hline
                
               Network& p & n & $\alpha$ & IME & IGE & $^{56}$Ni \\
                \hline
                \hline
               \net{}     & $2.4\times10^{-6}$& $2.6\times10^{-13}$ & $1.2\times10^{-4}$ & $5.9\times10^{-5}$ & $4.8\times10^{-2}$ & $6.4\times10^{-2}$ \\                
               \netnoec{} & $2.1\times10^{-4}$& $5.5\times10^{-14}$ & $1.1\times10^{-4}$ & $4.1\times10^{-4}$ & $1.4\times10^{-5}$ & $1.1\times10^{-1}$\\               
                \hline
    
       \end{tabular}
       \tablefoot{Yields of protons, neutrons, $\alpha$ particles, intermediate-mass elements, iron-group elements, and $^{56}$Ni in solar masses, for the 3D WD deflagration test at $t=2$~s (see Fig.~\ref{fig:COGroups} for a definition of these groups).}
       \label{tab:table_Defl_groups}
\end{table*}

\begin{figure}
\centerline{\includegraphics[width=0.5\textwidth]{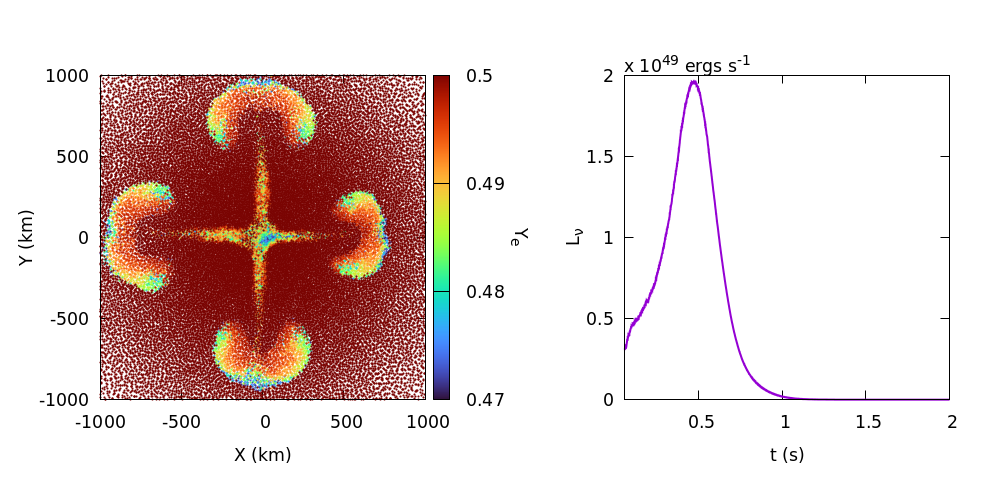}}
\hfil
\caption{Distribution of $Y_e$ in a XY slice (left) at $t=0.73$~s and evolution of neutrino luminosity in the deflagration test.}
\label{fig:3DYeneutriLum}
\end{figure}

\begin{figure}
\centerline{\includegraphics[width=0.5\textwidth]{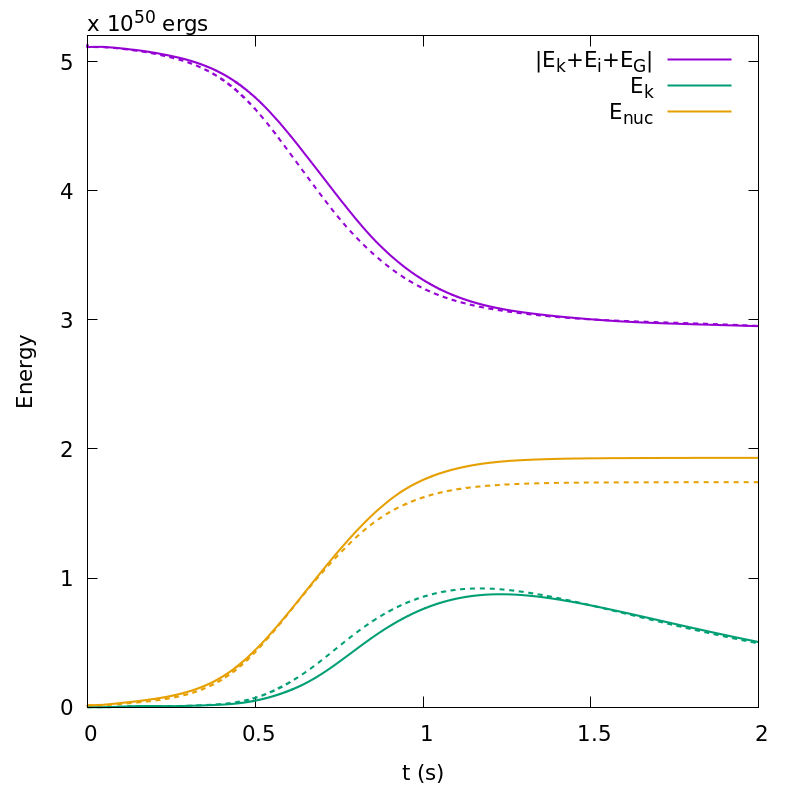}}
\hfil
\caption{Evolution of different energies during the deflagration of a WD. Solid lines correspond to \net{} and dashed lines to \netnoec{}.}
\label{fig:3Denergies}
\end{figure}

Figure~\ref{fig:3DTypeSNIanuc} shows the distribution of several nuclear species in an XY slice of the WD at two elapsed times. Four of the six high-temperature regions or hot bubbles are clearly visible in the slice. These hot bubbles with $T\simeq 8\times 10^9$~K rapidly ascend by Archimedian flotation and deform with the overall effect of increasing the contact surface between the hot ashes and the CO fuel, which accelerates the effective nuclear burning rate. After $t\simeq 1.2$~s the expansion of the WD quenches the nuclear combustion. The amount and distribution of the different families of isotopes in the burned regions are dependent on the \ele{} capture rate. Regions where \ele{} captures are relevant, last two columns in Fig.~\ref{fig:3DTypeSNIanuc}, produce more IGE elements and less $^{56}$Ni and vice-versa which agree with the results found in Sect. \ref{impactCh}. More detailed  values of the final yields of the different  families of elements are shown in Table~\ref{tab:table_Defl_groups}. Figure \ref{fig:3DYeneutriLum} depicts the distribution of the electron molar fraction $Y_e$ at $t= 0.73$ s (left panel) along the rising plumes of burnt material and with minimum values $Y_e\simeq 0.47$ at the tip of the bubbles and in the very central region of the WD. The released neutrinos  escaping from the WD produce a display, shown in the same figure (right panel) which could be detected if the event occurs at a sufficiently close distance. 

The evolution of different energies during the simulations with and without \ele{} captures is shown in Fig.~\ref{fig:3Denergies}. First note that the total amount of released nuclear energy, $E_{nuc}\simeq 1.8\times 10^{50}$ ergs, is not enough to unbind the star. The total amount of burnt material, $M_b\simeq 0.12$ M$_\sun$, is well below the limit of $\simeq 0.30$ M$_\sun$ needed to get  the WD  unbound \citep{bravo2009} and within the range of "failed deflagrations" described in \cite{ropke2007}. In contrast, the WD will undergo a pulsation having a second chance to explode at longer times, during the re-contraction period \citep{bravo2009b}. Interestingly and according to Figs.~\ref{fig:3Denergies} and \ref{fig:3DTypeSNIanuc}, the inclusion of \ele{} captures in the hydrodynamic calculation weakens the kinetic energy during the stages of fast combustion at $t\le 1.2$ s and reduces the vertical development  of the rising plumes of burnt material . Hence, we conclude that \ele{} captures  is a necessary ingredient to perform high-precision multi-D calculations of the explosion of a WD approaching the Chandrasekhar-mass limit.   

 The network \net{} is therefore capable to handle all nuclear combustion stages: initial explosive induction of the CO fuel, rapid combustion, QNSE and NSE and  breaking of the equilibrium during the fast expansion  in a multi-D simulation of the explosion of a massive WD. It also provides a reasonable first picture of the produced nucleosynthetic yields. The time-step always remained $\Delta t\simeq 2\times 10^{-4}$ s in the computed models above, an affordable time-step which  is not  very different from the current hydrodynamic Courant time.

\subsection{He burning}
\label{sec:helium_1}

Helium combustion might take place in the outer layers of an accreting white dwarf. Material from a companion star might be ripped away and fall onto the surface of the WD, where it piles up instead of burning. When density is high enough, an ignition will take place at one point (or several) of the He shell, triggering a detonation wave that propagates in all directions, eventually igniting the C+O core of the WD. This is the so-called double detonation mechanism (DDT) \citep{woosley1994}.

In this scenario, the densities in the He shell are significantly lower than those in the core of the WD, with $\rho_6 \sim 0.5-2$, and maximum temperatures of $T_9\sim 2.5-5$. With such conditions, electron captures are expected to play no significant role.

\subsubsection{Calculations with \net{} and \netalpha{}}

\begin{figure*}
\centerline{\includegraphics[width=\textwidth]{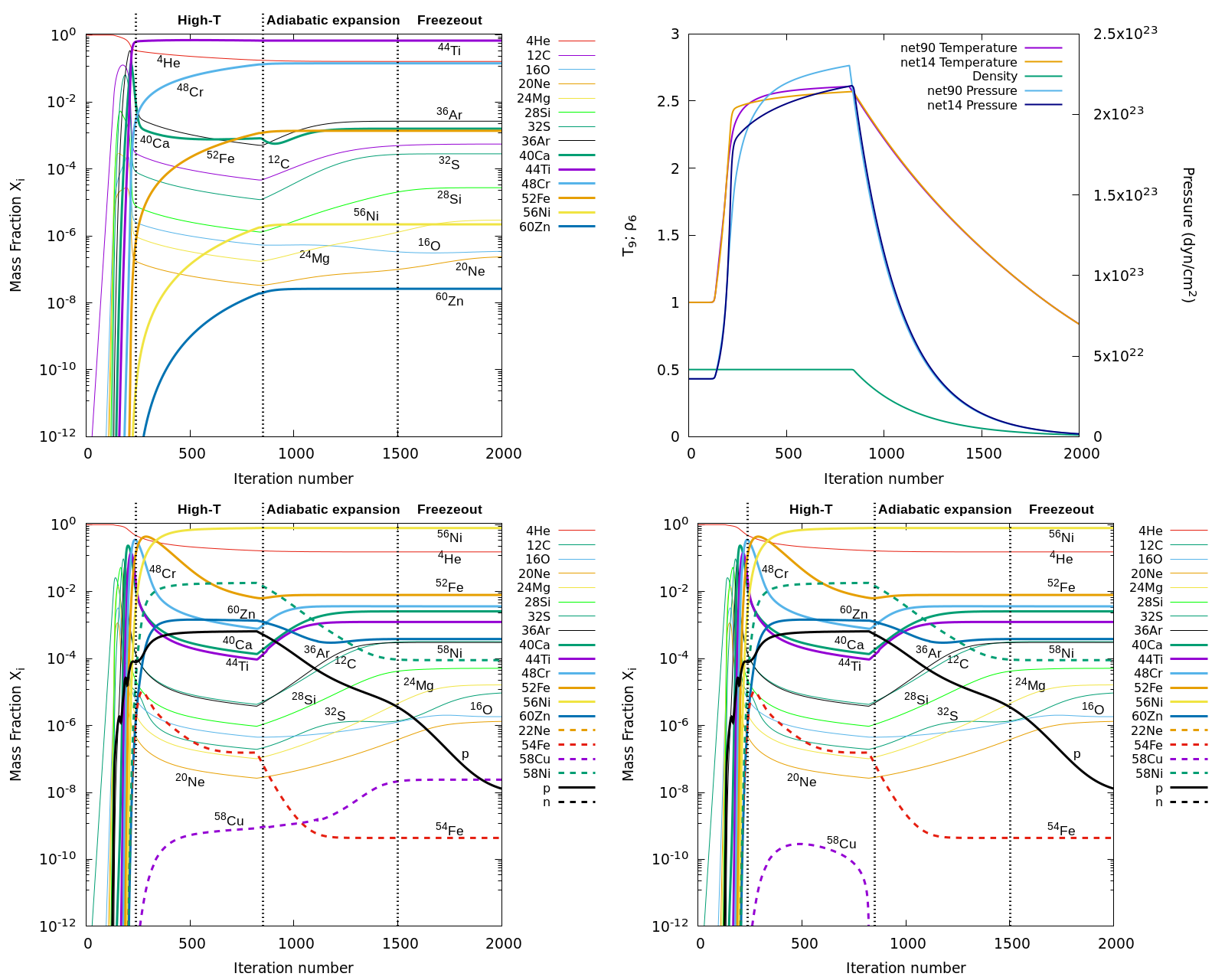}}
\hfil
\caption{Results of the He test. Evolution of mass fractions obtained with \emph{net14} (top left), \netnoec{} (bottom left), and \net{} (bottom right). Vertical dashed lines show the approximate limits of the high-temperature combustion region, when the artificial adiabatic expansion starts, and the region where nuclear reactions freezeout. The panel on the top right shows the evolution of temperature and pressure for \net{} and \netalpha{}. The calculation with \netnoec{} follows exactly the same trajectory as with \net{} and is not shown. The density evolution is the same for all three nuclear networks.}
\label{fig:Hetest}
\end{figure*}

Figure~\ref{fig:Hetest} shows the result of the He burning test using the initial conditions described in Table~\ref{tab:testsICs}. The top-left panel shows the outcome using \netalpha{}, while the lower panels present the results using \netnoec{} (left) and \net{} (right). When comparing all three results, it is evident that \emph{net14} provides very different yields, dominated by lighter metals $^{44}$Ti, $^{48}$Cr, and $^{36}$Ar. The larger networks products are dominated by elements of the iron family, such as  $^{56}$Ni and $^{52}$Fe. 
The reason for this is the architecture of the networks. The only way \netalpha{} can reach $^{56}$Ni is through alpha captures, which are very suppressed at the upper end of the network due to the low density and temperature. On the other hand, \netnoec{} (left) and \net{} (right) have many other paths through proton captures. These results imply that a small network such as \netalpha{} is not suitable for explosive scenarios with low densities in terms of the yields produced.

Furthermore, the thermodynamic evolution of \netalpha{} and \net{} also shows significant differences, as shown in the top-right panel of Fig.~\ref{fig:Hetest}. Temperature and pressure values during the high-temperature episode are lower than those obtained with \net{}, especially pressure. However, having a correct representation of pressure is crucial in the development of self-sustained helium detonations that trigger the explosion. Additionally, as expected at this low density, the inclusion of electron captures has no effect on either the temperature or the pressure. Only elements with very low abundance might be affected, such as $^{58}$Cu. Unlike the precedent tests, the electron mass fraction slightly increases $Y_e=0.5+1\times 10^{-8}$ at the freeze-out in \net{}, due to the positron captures on free neutrons, which are not blocked by degeneracy effects at the densities and temperatures involved in this scenario.


\subsubsection{Verifying \net{} with \netWN{}}
\label{sec:helium_2}

\begin{figure}
\centerline{\includegraphics[width=0.5\textwidth]{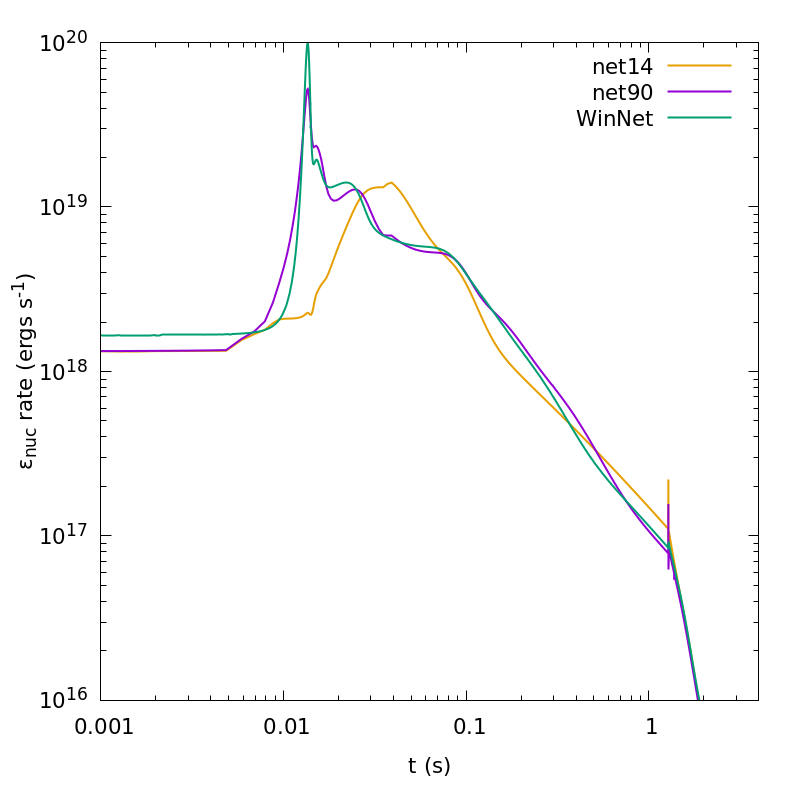}}
\hfil
\caption{Results of the He test. Evolution of the nuclear energy generation rate calculated with \netalpha{}, \net{}, and \netWN{}. }
\label{fig:Henucrate}
\end{figure}

We postprocessed the calculations of \net{} with both \netalpha{} and \netWN{} and analyzed and compared the ensuing results. Figure~\ref{fig:Henucrate} shows the evolution of the nuclear energy generation rate calculated with the three networks. The agreement between \net{} and \netWN{} is fairly good, both lines stay close but the larger network leads to a more pronounced, albeit narrow, maximum at $t\simeq 0.012$~s. The match of the reduced \netalpha{} is clearly worse, which reinforces the conclusion in Sect.~\ref{sec:helium_1} that reduced $\alpha-$like networks are not adequate to address Sub-Chandrasekhar-mass explosion models of Type Ia supernova.

\begin{figure}
\centerline{\includegraphics[width=0.5\textwidth]{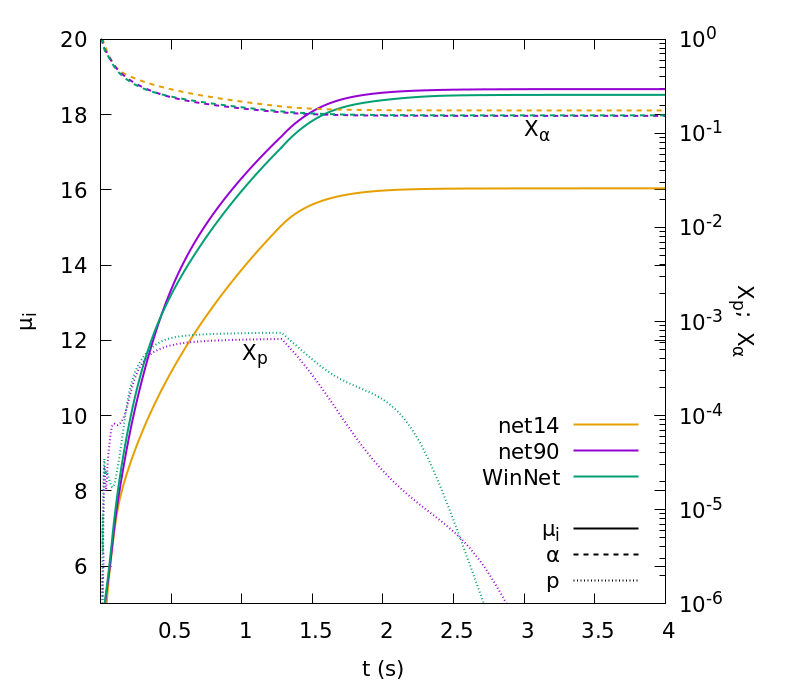}}
\hfil
\caption{Results of the He test. Evolution of the mean molecular weight of ions, $\mu_i$ (leftmost Y-axis), and protons and helium mass fractions (rightmost y-axis) calculated with \netalpha{}, \net{}, and \netWN{}. }
\label{fig:HeXaXp}
\end{figure}

 The mean molecular weight of ions calculated with \net{} perfectly fits that of \netWN{} while \netalpha{} leads to somewhat lower values (see Fig.~\ref{fig:HeXaXp}). The corresponding value of $\alpha$ particles is shown in the same figure. Free protons evolve similarly until $t\simeq 1.5$~s but the \net{} estimate remains below that of \netWN{} at $t\ge 1.5$, due to the increasing number of p-production channels in the larger network at later times.   

\subsection{Si burning}
The photodesintegration of Si is relevant in scenarios such as a core-collapse supernova, where the core of a massive star implodes due to its own mass and subsequent deleptonization \citep[e.g.,][]{Woosley2002,Janka2007}. The temperature rises and heavy elements are photodissociated. This process happens nearly isothermally because of the large thermal conductivity of the electrons. To mimic such conditions, we artificially increased the specific heat, which is usually provided by the EOS, so that $\Delta T\sim 0$. Table~\ref{tab:testsICs} provides the initial conditions and Fig.~\ref{fig:Sitest} shows the results of the test. The setting of this test is similar to that described in \cite{Timmes2000} who compared the performance between two reduced nuclear networks in light of the results obtained with a larger network with N=489 isotopes.

\subsubsection{Calculations with \net{} and \netalpha{}}

\begin{figure*}
\centerline{\includegraphics[width=\textwidth]{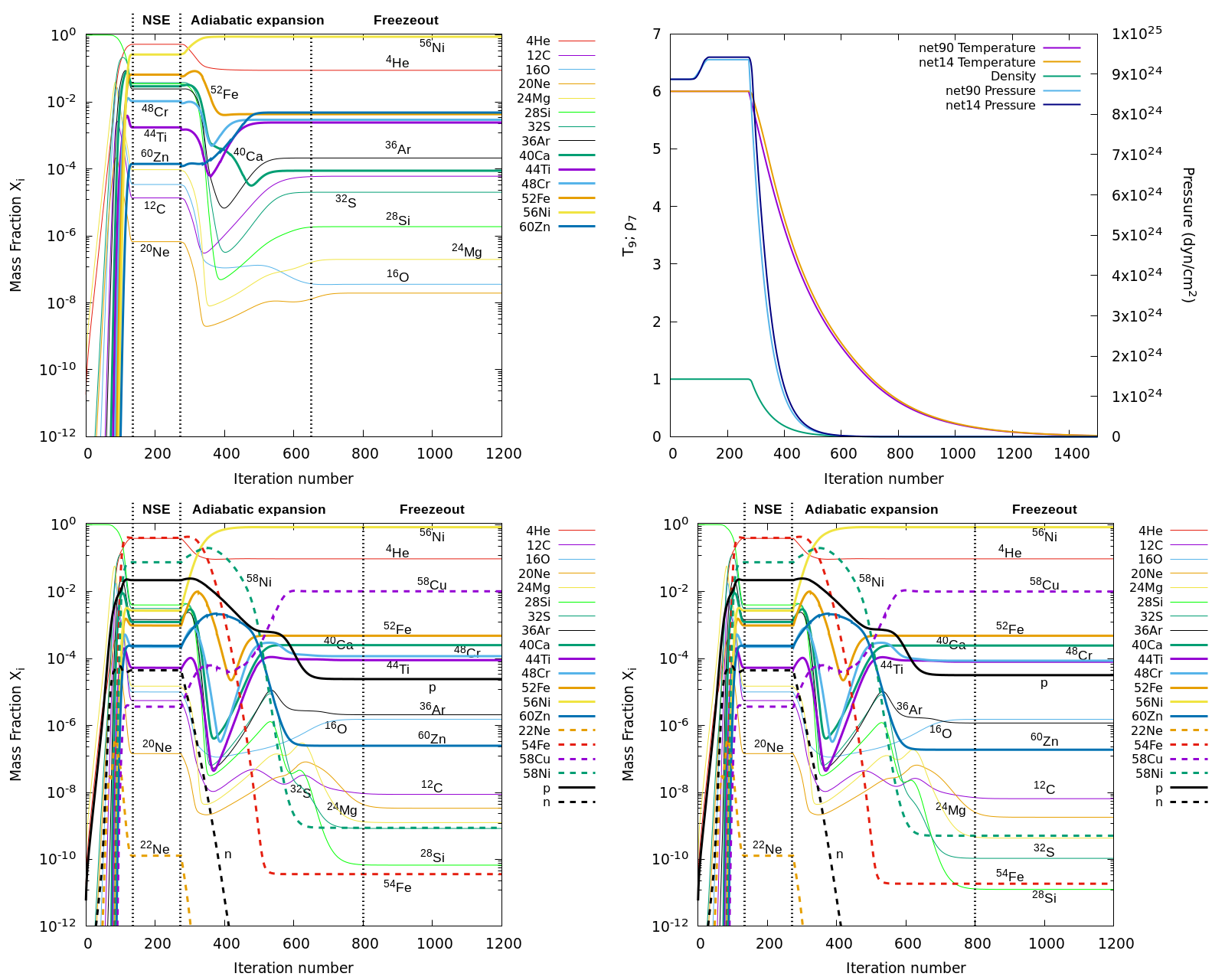}}
\hfil
\caption{Results of the Si test. Evolution of nuclear mass fractions obtained with \emph{net14} (top-left), \netnoec{} (bottom-left), and \net{} (bottom-right). Vertical dashed lines show the approximate limits of the NSE, when the artificial adiabatic expansion starts, and the region where nuclear reactions freezeout. The panel in the top right shows the evolution of temperature and pressure for \netalpha{} and \net{}. The evolution of density is the same for all three nuclear networks.}
\label{fig:Sitest}
\end{figure*}

The top-left panel of Fig.~\ref{fig:Sitest}  clearly shows that, with this high temperature, Si is rapidly destroyed, producing a high amount of alpha particles. These are used to produce all the elements of the alpha-chain, reaching NSE very quickly. In particular, $^{56}$Ni and alpha particles are the most abundant elements. During adiabatic expansion, a large fraction of this reservoir of alpha particles is used to climb the alpha chain using the seeds present at NSE, producing even more $^{56}$Ni and an overall decrease in all the other elements. Once the temperature has decreased sufficiently ($T_9 \lesssim 5$), no more $^{56}$ Ni is produced, leading to a slight increase in all other elements through alpha captures until they slowly quench as reaction channels are closed from heavier to lighter elements. The final result after freezeout is dominated by $^{56}$Ni and $^4$He, followed by heavy elements and finally lighter species.

The bottom panels of Fig.~\ref{fig:Sitest} show the same test performed with \netnoec{} (left) and \net{} (right). First, we notice that there are very small differences among them. This is due to the low initial density of this test ($\rho_9=0.01$). At this density electron captures are subdominant, and only slight differences can be noticed in elements with very low mass fraction. The inclusion of proton and neutron captures is what has the greatest impact compared to the results obtained with \netalpha{}. In \netnoec{} and \net{} there are many more nuclear paths open for the burning of $^{28}$Si. This can be seen in a much lower mass fraction of $^{28}$Si in the NSE, the dominance of neutronized isotopes, such as $^{58}$Ni and $^{54}$Fe, and the high abundance of protons in that regime.

\begin{figure}
\centerline{\includegraphics[width=0.5\textwidth]{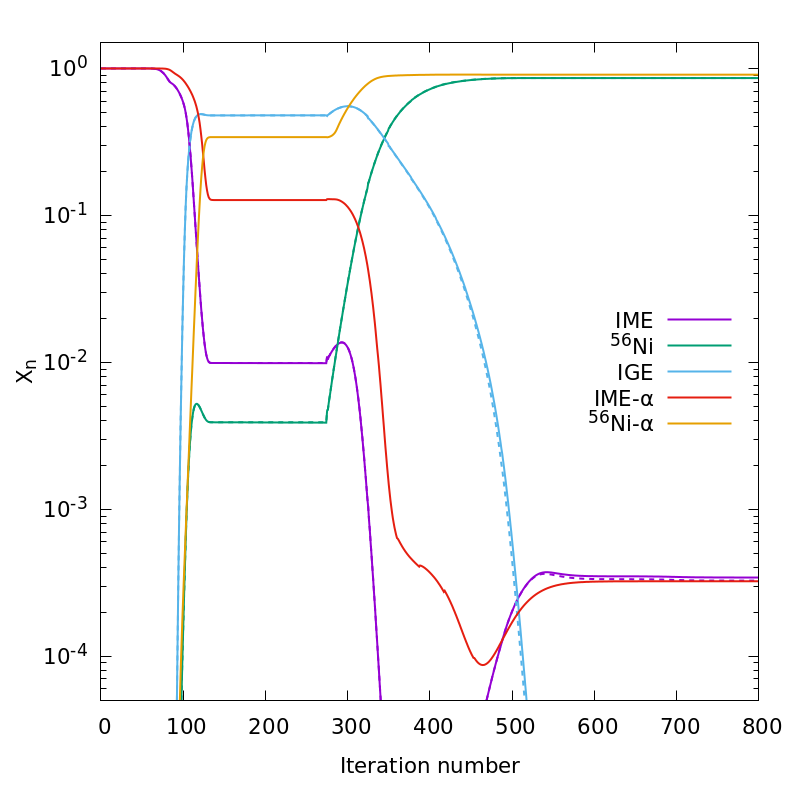}}
\hfil
\caption{Results of the Si test. The mass fractions in Fig.~\ref{fig:Sitest} are grouped in three nuclear families (see Fig.~\ref{fig:COGroups} for a definition of these groups) for \netnoec{} (dashed lines), \net{} (continuum lines) and the \netalpha{} ($\alpha$-network).}
\label{fig:SiGroups}
\end{figure}

Once the adiabatic expansion sets in the dominant isotopes during the NSE, $^{58}$Ni and $^{54}$Fe have their mass fractions severely reduced. The mass fraction of alpha particles also decreases at the beginning of the expansion, facilitating the build up of $^{56}$Ni. 
The freezeout is dominated by $^{56}$Ni and $^4$He, followed with a stratification similar to that of \netalpha{} but with more abundance of heavier elements.
From the top-right panel we can see that the temperature and pressure evolution is very similar in all cases, and that the inclusion of electrons has no noticeable effect. Figure~\ref{fig:SiGroups} shows the mass fractions of IME, IGE, and $^{56}$Ni-like elements. In contrast to the CO test mentioned earlier, the inclusion of \ele{} captures on protons does not significantly affect the results. Thus, \netnoec{} and \net{} virtually produce the same results. However, the differences between these and \netalpha{} are significant during the isothermal combustion phase, with the reduced network generating a substantially higher amount of IME and $^{56}$Ni. Nevertheless, all networks synthesize similar final amounts of nickel and IME.

\subsubsection{ Verifying \net{} with \netWN{}}

\begin{figure*}
\sidecaption
\includegraphics[width=12cm]{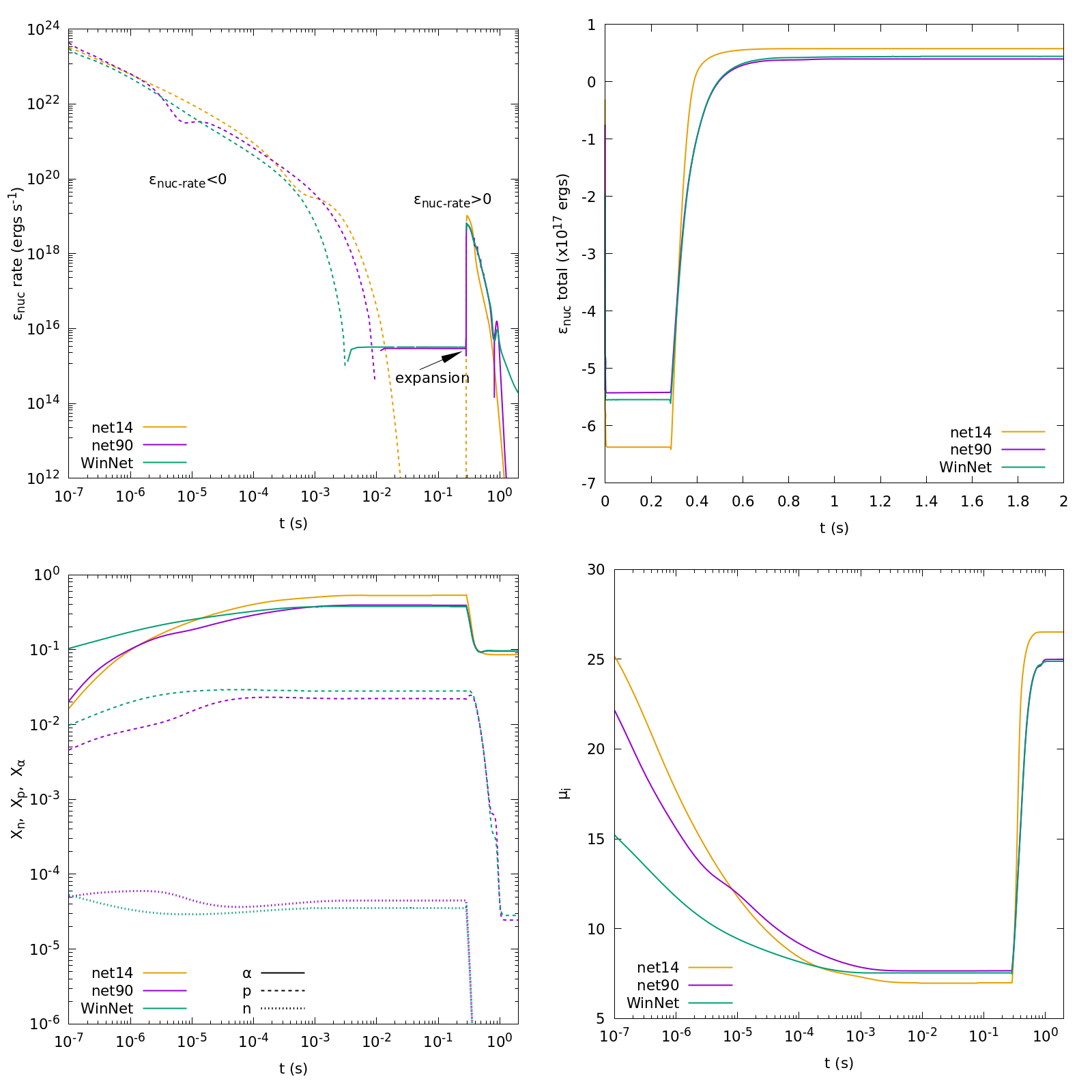}
\hfil
\caption{Results of the Si test. The upper panels show the evolution of the nuclear energy generation rate (left) and total released nuclear energy (right), while the lower panels depict the evolution of the mass fraction of the lighter elements $n$, $p$, and $\alpha$ particles (left), and mean molecular weight of ions, $\mu_i$ (right)}
\label{fig:Sitestnet89comparison}
\end{figure*}

As in the precedent tests, we take advantage of the temperature and density evolution obtained with \net{} to 'post-process' the calculation with \netalpha{} and \netWN{} and obtain the nucleosynthetic yields and release of nuclear energy, which are in turn used to verify the reduced networks.  The main results regarding these magnitudes are shown in Figure \ref{fig:Sitestnet89comparison}. The upper-left panel depicts the nuclear energy generation rate that proceeds in two separate regimes. For $t\lesssim 0.01$~s nuclear reactions are endoenergetic due to the prevalence of photodisintegration reactions. In this region \netalpha{} and \net{} match better with each other than with \netWN{}, although overall \net{} evolves closer to \netWN{} than \netalpha{}. The evolution in this region agrees with that obtained by \cite{Timmes2000}, although their large-network calculation stays closer to their reduced network estimations than in our work. We observe slight differences between the networks, as (n,p) reactions are dominating for the prevailing conditions. Although all networks strive to achieve an NSE composition, \netWN{} can achieve it fastest as, in contrast to the other networks, it includes (n,p) reactions. At $t\ge 0.01$~s the energy release is positive and the match between \net{} and \netWN{} improves in that region while \netalpha{} fails to reproduce $\dot\epsilon_n$ between $0.01\le t \le 0.3$ s. The upper-right panel shows the evolution of the nuclear energy released until the elapsed time $t$, $\epsilon_n(t)$ on a linear scale. Although the pure $\alpha-$ chain gives in general poorer results than \net{}, the error in the released energy at freeze-out of both networks is similar (see also the second to last column in Table~\ref{tab:table_Si})    

The lower panels of Fig.~\ref{fig:Sitestnet89comparison} show the evolution of the mass fraction of light elements and the mean molecular weight of nuclei calculated with the three networks. The calculation with \netWN{} leads to a larger population of $\alpha$ particles for $t\lesssim 10^{-3}$~s which lowers the value of $\mu_i$. However, \net{} and \netWN{} give almost identical results when $t\ge 10^{-3}$~s. Overall, the values of $\alpha$ and $\mu_i$ obtained with \netalpha{} do not compare so well with \netWN{} as those computed with \net{}. The evolution of free neutrons and protons obtained with \net{} and \netWN{} is fairly similar.

Additional quantitative information on the Si test is included in Table~\ref{tab:table_Si}. The most abundant element when nuclear reactions have stopped is $^{56}$Ni, which is abundantly produced and with similar values in all three cases (third column). The value of $Y_e$ remains very close to $Y_e=0.5$ in all calculations (fifth column) and consequently, the energy drained by the freely escaping neutrinos $E_\nu$ is only a small fraction of the total released energy at the freeze-out time (last column). However, $E_\nu$ is more than ten times larger in the \netWN{} calculation because \ele{} captures on nuclei proceed from the very beginning while \net{} has to wait until there are enough protons in the mixture.    

\begin{table*}
        \centering
        \caption{Results of the Si test.} 
       \begin{tabular}{cccccccc} 
               \hline
                
               Network& $X_{\alpha}$& $X_{max}$& $\mu_i$&$ Y_e$ &  $\epsilon_{n}$&$E_{\nu}$& \\
               &&& && $\times 10^{16}$ ergs$\cdot$g$^{-1}$ & $\times 10^{15}$~ergs$\cdot$g$^{-1}$&  \\
                \hline
                \hline
\netalpha{} & $0.085$& $0.900$ ($^{56}$Ni)  & $26.51$ &$0.50000$& $5.80$& $ - $\\                
               \net{}  & $0.096$& $0.843$ ($^{56}$Ni)  & $24.98$ &$0.49994$& $3.98$& $0.13$\\
                \netWN{} & $0.096$& $0.853$ ($^{56}$Ni)& $24.88$&$0.49979$& $4.43$& $1.9$\\
                
                \hline
    
       \end{tabular}
       \tablefoot{Comparison among several magnitudes calculated with \netalpha{}, \net{}, and \netWN{} at $t=2$~s, when complete freeze-out of the reactions has been achieved. Columns are: the name of the nuclear network, mass fractions of $\alpha$ particles and more abundant nuclei, mean molecular weight of ions, electron abundance, total released nuclear energy, and energy carried out by the freely escaping neutrinos, respectively.}
       \label{tab:table_Si}
\end{table*}

\section{Computational cost}
\label{sec:performance}
In previous sections we showed that using a medium-sized nuclear network coupled with hydrodynamical simulations has clear advantages in terms of accuracy and reliability of the results compared to small, $\alpha$-like, networks. However, this comes with a cost. In this section, we present the measurements of the computational cost of each nuclear network. 

Table ~\ref{tab-1} shows the average computational performance of \netalpha{}, \netnoec{}, and \net{} for 100 full runs of the CO test executed with a single thread on an Intel i7-4790 CPU (3.6~GHz), compiled with \verb|gfortran| and \verb|-O3| flag. All runs execute $\sim1500$ global iterations and $\sim6000$ NR iterations, which correspond to $\sim600,000$ calls to each nuclear subroutine. The columns of Table~\ref{tab-1} show, from left to right, the name of the network, the number of global iterations, the ratio of NR iterations per global iteration, the average time per global iteration (that is, for the entire cycle of NR iterations), the average time per network call (namely, per individual NR iteration), the average percentage of time spent on the sparse matrix solver, and the scaling factor for each network call with respect to \netalpha{}.

From these results, the computational cost of the network scales approximately as $N\log N$ with the number of reaction rates (being $N=17$ for \netalpha{} and $N=161$ for \net{}). The overload of including $e^-$ and $e^+$ captures is very small ($\sim0.4\%$). Approximately a fourth of the time per network call is spent in the sparse matrix solver. Hence, any improvement in this part of the code would have a noticeable effect overall. 

Coupling chemical equations with temperature increases the algebraic complexity of the integration scheme, and one may wonder if the increment in the timestep during QNSE-NSE conditions compensates for the extra algebraic effort. At $\rho_9=1, T_9=8$ the timestep during the NSE plateau calculated with \net{} is $\Delta t\simeq 10^{-4}$ s, not much different than the Courant time at the center of the WD in current multi-D calculations $\tau_C= 0.4~(\Delta x/c_s)\simeq 4\times 10^{-4}$~s for a cell size $\Delta x=5$ km and sound speed $c_s=5000$~km$\cdot$s$^{-1}$. Removing the thermal implicit coupling in \net{} drops the time-step to $\Delta t\simeq 10^{-8}-10^{-10}$~s to control the numerical fluctuations in temperature and abundances\footnote{Such tiny number is of the order of the characteristic nuclear reaction time $\tau_n\simeq -Y_i/\dot Y_i = (Y_j r_{ij})^{-1}$ affecting the equilibrium value of relevant NSE nuclei as for example $^{56}$Ni in the reaction $p+^{56}\text{Ni}\leftrightarrow ^{57}\text{Cu}+\gamma$. At $\rho_9=1$ and $T_9=8$, turning off \ele{} captures and with $Y_p$ and $Y_{^{56}\text{Ni}}$ taken from \net{}, it gives $\tau_n\simeq 6\times 10^{-9}$~s.}. Therefore, incorporating temperature into the Jacobian solver of the nuclear network increases the timestep by a factor $10^{4-5}$ when $T_9 \ge 8$.   

A different stabilizing procedure, described in \cite{Paxton2015}, consists on raising  the floating point limit in arithmetic operations, for example from the commonly used  double precision to quadruple precision, which increases the stability of the integration. 
Both approaches are not exclusive and could be used jointly.

\begin{table*}
\centering
\caption{Computational cost of each nuclear network.}
\label{tab-1}       

\begin{tabular}{ccccccc} 
\hline
        \toprule
         \multirow{2}{*}{Network} & \multirow{2}{*}{It.} & \multirow{2}{*}{NR/It.} & {$\overline{t}$/It.} &{$\overline{t}$/NR} & $\overline{t}_{\mathrm{matrix}}$ & \multirow{2}{*}{scaling} \\
         &&&($\times 10^{-5}$~s)&($\times 10^{-6}$~s)&(\%)&\\
         \hline
         \hline
          \netalpha{} & 1501 & 3.47 & $1.11\pm0.23$  & $3.21\pm0.67$  & $12.35\pm8.18$ & -\\
          \netnoec{}  & 1524 & 3.72 & $17.85\pm0.62$ & $47.93\pm1.67$ & $25.02\pm2.45$ & $\times14.93$\\
          \net{}      & 1508 & 3.85 & $18.53\pm0.66$ & $48.10\pm1.72$ & $25.07\pm2.57$ & $\times14.98$\\
          \hline
   \end{tabular}
   \tablefoot{All data refer to the full CO test averaged over 100 runs. Columns show the name of the network, the number of global time iterations, the number of NR iterations per global time iteration, the average wall-clock time per time iteration, the average wall-clock time per NR iteration, the percentage of wall-clock time spent on the sparse matrix solver, and the computational cost per network call with respect to \netalpha{} (ratio among the pertinent rows in the fifth column). The errors correspond to the standard deviation.}
\end{table*}

\section{Conclusions}
\label{sec:conclusions}

In this work, we propose and verify a medium-sized nuclear network especially addressed to multidimensional simulations of Type Ia supernova explosions. The network \net{}, with 90 species from $^4$He to $^{60}$Zn has two distinctive features. First, the integration of nuclear species is carried out using an implicit scheme that also includes temperature in the Jacobian matrix \citep{mueller86, cab04}. Such thermal coupling stabilizes the integration scheme during the QNSE and NSE stages, allowing to take considerably larger timesteps than schemes where temperature changes are calculated in an explicit form. Furthermore, both the implicit integrator and the matrix solver have been improved with respect to those described in paper~I. 

The second feature, and a novelty of this work, is the inclusion of weak interaction processes in the reduced network. Electron captures on protons and nuclei have been traditionally incorporated into large nuclear networks and NSE tables supporting spherically symmetric explosion models of SNIa \citep[e.g.,][]{nom84} and CCSN and in a few multi-D calculations of Type Ia supernova explosions  that interpolate the value of $Y_e$ from large NSE tables \citep[e.g.,][]{moll2013}. Weak processes are also currently incorporated to post-process the output of multi-D calculations of both kinds of supernova explosions. Nevertheless, and as far as we know,  these processes have been ignored by all reduced networks devoted to multi-D hydrodynamic simulations so far. In particular, electron and positron captures on protons and neutrons, respectively. Although it may seem that including weak processes only on free particles is too simplistic, we have shown that \net{} manages to account for $\simeq 2/3$ of electron captures under the typical conditions prevailing in Type Ia supernova explosions. Therefore, the pressure of electrons is substantially better estimated than if such reactions are ignored, allowing for a more accurate depiction of the dynamics of the explosion. 

We verified \net{} by direct comparison with the results obtained with a much larger network, \netWN{} \citep{reichert2023}, under typical supernova conditions. These include explosive carbon and oxygen mixtures with maximum densities, prior expansion, in the range $10^8$\dens{}$\le \rho_0\le 5\times 10^{9}$\dens{} and edge-lit helium ignition at $\rho\simeq 10^6$\dens{}. These density ranges are representative of many SNIa explosion models currently considered in the literature, either of Chandrasekhar or Sub-Chandrasekhar mass type.

We found that in the high density regime \net{} gives a rather good approach to the released nuclear energy, total pressure and mean molecular weight of ions, $\mu_i$, and provides a much better depiction of produced nuclear yields than those obtained with small networks such as \netalpha{}. According to Table~\ref{tab:table_W7}, the medium-sized network calculation gives a good estimate of $\mu_i$ and total released nuclear energy of a toy model that mimics the central shell of the W7 supernova model by \cite{nom84} at freeze-out time. The relative deviations of both magnitudes with respect to those computed with \netWN{} are smaller than $4\%$. However, the deviations of the nuclear energy generation rate could occasionally be larger, as shown in Figs.~\ref{fig:COenermuitest} and \ref{fig:testCOnuctime}. Table~\ref{tab:table_W7} also shows that \net{} leaves more uncombined $\alpha$-particles, which is attributable to the progressive loss of $\alpha$-reaction channels as neutronization increases. 

Explosive helium ignition at $\rho\simeq 10^6$\dens{} is also supported by \net{}. This corresponds to point-like edge-lit ignitions near the surface of WDs with masses $0.6\le M_{wd}\le 1.0$~\msun{}, propagating as detonation waves \citep{gronow21}. Too small networks, such as \netalpha{}, are not very adequate here because temperature and pressure jumps are not accurately captured. This is shown in the upper-right panel of Fig.~\ref{fig:Hetest}. In contrast, the results with \net{} are fairly in agreement with those with \netWN{} in terms of nuclear energy generation rate $\dot\epsilon_n$, $\mu_i$, dominant nuclei, and mass fractions of light particles (Figs.~\ref{fig:Henucrate} and \ref{fig:HeXaXp}) which ensures pressure compatibility. Electron and positron captures do not play a significant role in this density regime. 

Finally, we have applied \net{} to the combustion of silicon at densities 
$\rho_0\simeq 10^7$\dens{}, this being a standard nuclear test \citep[e.g.][]{Timmes2000} that is more akin to Core Collapse Supernova (CCSN) than to SNIa. In this test, the silicon is rapidly disintegrated by photons, leading to a negative generation of nuclear energy during some time. According to Fig.~\ref{fig:Sitestnet89comparison}, the late stages of this phase are neither well described by \netalpha{} nor by \net{}, although the latter is slightly closer to the \netWN{} solution. After a while, the nuclear generation rate turns to positive values and the match between \net{} and \netWN{} improves, while the reduced \netalpha{} is still unable to give realistic results. The electron abundance does not change too much during the interval of time considered in the test. However, both the change of $Y_e$ and the energy released by neutrinos are much larger when calculated with \netWN{}. All of this suggests that, in agreement with \cite{navo2023}, caution should be taken when using medium-sized networks for CCSN related scenarios.  

In summary, we proposed and verified a nuclear network implicitly coupled with temperature and built around a standard $\alpha$-chain but expanded to host 90 nuclei including free protons, neutrons, and electrons (Fig.~\ref{fig:arquitecture_net90}). This network includes \ele{}+p and \pos{}+n capture processes, which make it ideally suited for being implemented in multidimensional simulations of Type Ia Supernova. The feasibility of net90  to  handle  the deflagration of a massive  WD  in a three-dimensional calculation  was   confirmed in Sect. \ref{sec:impactCh3D}.

The price to pay for such a better depiction of the nuclear energy generation rate and the effect on the overall physical state of matter comes in the form of a larger computational cost. This cost scales roughly as $N\log N$, with the number of nuclear reactions implemented. According to our measurements, \net{} is a factor $\times15$ slower than \netalpha{}. Even if such a factor cannot change, there is still room for future improvements that could be made to reduce the overall wall-clock time per network call, for example in the architecture of the matrix solver which could be of sparse type, such as the widely used \verb|PARDISO|, or dense matrix solvers such as \verb|LAPACK|. Admittedly, an increased computational cost is unavoidable when using a larger nuclear network. Yet, the benefits of considerably more accurate simulations outweigh this cost.
 
\begin{acknowledgements}
We thank the anonymous referee for valuable comments and suggestions that helped to improve this manuscript. This work has been supported by the Spanish MINECO grant PID2020-117252GB-100 and by the AGAUR/Generalitat de Catalunya grant SGR-386/2021 (DGS). It has also been supported by the Swiss Platform for Advanced Scientific Computing (PASC) project "SPH-EXA: Optimizing Smoothed Particle Hydrodynamics for Exascale Computing" (RC). The authors acknowledge the support of the Center for Scientific Computing (sciCORE) (\url{https://scicore.unibas.ch/}) at the University of Basel, where part of these calculations were performed. MR acknowledges support from the Juan de la Cierva program (FJC2021-046688-I) and the grant PID2021-127495NB-I00, funded by MCIN/AEI/10.13039/501100011033 and by the European Union "NextGenerationEU" as well as “ESF Investing in your future”. Additionally, he acknowledges support from the Astrophysics and High Energy Physics program of the Generalitat Valenciana ASFAE/2022/026 funded by MCIN and the European Union NextGenerationEU (PRTR-C17.I1) as well as support from the Prometeo excellence program grant CIPROM/2022/13 funded by the Generalitat Valenciana. AP acknowledges support from the U.S. Department of Energy, Office of Science, Office of Nuclear Physics, Award Number DE-SC0017799 and Contract Nos. DE-FG02-97ER41033 and DE-FG02-97ER41042. AA was supported by the Deutsche Forschungsgemeinschaft (DFG, German Research Foundation) -- Project-ID 279384907 - SFB 1245 and the State of Hessen within the Research Cluster ELEMENTS (Project ID 500/10.006). Finally, the support of the European COST Action CA16117 Chemical Elements as Tracers of the Evolution of the Cosmos (ChETEC) is acknowledged.
\end{acknowledgements}

\bibliographystyle{aa}
\bibliography{bibliography_r}

\end{document}